\documentclass[fleqn,usenatbib]{mnras}

\usepackage{graphicx}	
\usepackage{amsmath}	
\usepackage{amssymb}	
\usepackage{multicol}        
\usepackage{bm}		
\usepackage{pdflscape}	
\usepackage{dcolumn}
\usepackage{xcolor}
\usepackage[T1]{fontenc}
\usepackage{ae,aecompl}
\usepackage{hyperref}
\usepackage{natbib}
\usepackage[normalem]{ulem}
\usepackage{fontawesome}
\usepackage{newtxtext}

\newcommand{\redmagic}{\texttt{redMaGiC} }
\title[Lagrangian bias emulator]{The cosmology dependence of galaxy clustering and lensing from a hybrid $N$-body--perturbation theory model}

\author[Kokron et al.]{Nickolas Kokron$^{1,2}$
\thanks{Contact e-mail: \href{mailto:kokron@stanford.edu}{kokron@stanford.edu}}%
, Joseph DeRose$^{3,4}$, Shi-Fan Chen$^{3}$, Martin White$^{3,5}$, Risa H. Wechsler$^{1,2}$%
\\
$^{1}$ Kavli Institute for Particle Astrophysics and Cosmology and Department of Physics, Stanford University, 382 Via Pueblo Mall, Stanford, CA 94305, USA \\ 
$^{2}$ Kavli Institute for Particle Astrophysics and Cosmology, SLAC National Accelerator Laboratory, 2575 Sand Hill Road, Menlo Park, CA 94025, USA \\
$^{3}$ Department of Physics, University of California, Berkeley, 366 LeConte Hall, Berkeley, CA 94720, USA\\
$^{4}$ Santa Cruz Institute for Particle Physics, University of California, Santa Cruz, CA 95064, USA \\
$^{5}$ Lawrence Berkeley National Laboratory, 1 Cyclotron Road, Berkeley, CA 93720, USA \\
}
\pubyear{2021}
\date{}

\begin{document}

\label{firstpage}
\pagerange{\pageref{firstpage}--\pageref{lastpage}}
\maketitle

\begin{abstract}
We implement a model for the two-point statistics of biased tracers that combines dark matter dynamics from $N$-body simulations with an analytic Lagrangian bias expansion. Using \texttt{Aemulus}, a suite of $N$-body simulations built for emulation of cosmological observables, we emulate the cosmology dependence of these nonlinear spectra from redshifts $z = 0$ to $z=2$. We quantify the accuracy of our emulation procedure, which is sub-per cent at $k=1\, h {\rm Mpc}^{-1}$ for the redshifts probed by upcoming surveys and improves at higher redshifts.  We demonstrate its ability to describe the statistics of complex tracer samples, including those with assembly bias and baryonic effects, reliably fitting the clustering and lensing statistics of such samples at redshift $z\simeq 0.4$ to scales of $k_{\rm max} \approx 0.6\, h\mathrm{Mpc}^{-1}$. We show that the emulator can be used for unbiased cosmological parameter inference in simulated joint clustering and galaxy--galaxy lensing analyses with data drawn from an independent $N$-body simulation.
These results indicate that our emulator is a promising tool that can be readily applied to the analysis of current and upcoming datasets from galaxy surveys.
\end{abstract}
\maketitle

\section{Introduction}
\label{sec:intro}

We are entering a golden era for studying the large-scale structure of the Universe. Over the next decade, ambitious imaging surveys will map out large swathes of the sky to unprecedented depths, imaging billions of galaxies and their shapes \citep{Ivezic:2008fe, laureijs2011euclid, dore2015cosmology, dore2019wfirst}, enabling studies of weak gravitational lensing by the intervening distribution of matter \citep{Bartelmann:1999yn, Mandelbaum:2017jpr}. Weak lensing has only recently begun to contribute competitive cosmological constraints on dark matter and dark energy \citep{Abbott:2017wau, 2020arXiv200715632H}, but is one of the most promising future directions to pursue. Meanwhile, spectroscopic surveys will observe tens of millions of radial positions of galaxies \citep{2014PASJ...66R...1T, Aghamousa:2016zmz}, enabling unparalleled understanding of the spatial distribution of galaxies in our Universe. The cross-correlation between positions and lensing, galaxy--galaxy lensing, is and will continue to be a key driver of cosmological constraints from galaxy surveys.\par

The quality and quantity of these upcoming datasets imposes a significant challenge in their analysis. Even now, models for summary statistics such as correlation functions and power spectra are inadequate across the full range of scales probed by such surveys \citep{Krause:2017ekm, Nishimichi:2020tvu}. Either a large amount of the data must be discarded, or mitigation schemes must be developed to prevent contamination from scales where the models are insufficiently calibrated or constrained \citep{MacCrann:2019ntb, Park:2020ojp}. Models for clustering and lensing must be substantially improved if we are to extract the maximal information about the Universe we live in, from surveys that are already ongoing or planned. To date, two separate approaches have been developed to build models for the observables of cosmic surveys: analytically, through perturbative techniques, or numerically, using non-linear $N$-body simulations.\par

Perturbation theory provides a systematic, analytic way to compute $N$-point summary statistics to systematically higher precision and smaller scales \citep{Bernardeau:2001qr}.  Below the nonlinear scale the effects of these nonlinearities can be tamed and parametrized within the framework of effective theories \citep{Bauman12,Carrasco:2012cv,Vlah15}. This increased precision, however, comes at the cost of very large inaccuracies beyond the nonlinear scale at which the self-gravitating dark matter fluid ceases to be perturbative \citep{Blas14,McQuinn16}. In addition, perturbative frameworks provide a rigorous, first-principles approach to include physics beyond the standard $\Lambda$CDM model in large-scale structure observables such as neutrinos, baryonic effects and more exotic early-universe scenarios \citep{Lewandowski:2014rca,Senatore17,Aviles2020a,Chen:2020ckc, Lague:2020htq,Ivanov:2020ril, 2020arXiv200612420D,Aviles2020b}. Understanding the domain of applicability of perturbation theory is still an active field of research \citep{Baldauf:2015zga, Nishimichi:2020tvu,chen2020redshiftspace}. \par 
The other approach, simulation-based modelling, involves numerically solving the equations of motion for an initial distribution of matter \citep{hockney1988computer,Bagla:2004au,Kuhlen:2012ft}. The resulting catalogs can be analysed in a way analogous to data to obtain predictions of cosmological observables across a wide range of scales at the cosmological parameters of the simulation. \par 
However, a limiting factor in simulation-based analyses is that $N$-body simulations require significant
computational resources for a single realization. Thus, standard inference procedures such as Markov Chain Monte Carlo (MCMC) become prohibitively expensive when using models derived from simulations. \par 
In order to ameliorate the issues with simulation-based inference, recent developments in statistical learning have popularized so-called emulators as models \citep{Heitmann:2008eq,Heitmann:2009cu,Lawrence_2010}. Emulators combine a set of simulations that representatively sample cosmological parameter space with sophisticated regression techniques to `fill in the blanks' across parameter space. Once trained, an emulator provides rapid evaluations of a model which can be seamlessly integrated in analysis pipelines. For example, recent emulators for the nonlinear matter power spectrum \citep{Knabenhans:2018cng} have runtimes with negligible overhead compared to the underlying Boltzmann codes used for linear predictions. \par
While galaxy surveys observe luminous tracers of the underlying dark matter density distribution, most suites of $N$-body simulations used to construct emulators deal only with the dark matter component. Thus, emulators for galaxy survey observables are presented with the additional challenge of capturing the relationship between the galaxy distribution and the underlying dark matter. Understanding the details of this relationship, known as the galaxy--halo connection, is an active field of research \citep[see e.g.][for a recent review]{Wechsler:2018pic}. Even for well-studied samples of galaxies, there are no consensus models to describe this relationship. For any given model of the galaxy--halo connection, an entirely new emulator has to be trained \citep{Kwan:2013jva,Wibking:2017slg,Zhai:2018plk,mclaughlinssemblybias}. Emulation of models with a large number of free parameters is also a challenging task, with techniques such as Gaussian processes scaling as $\mathcal{O}(N^3)$ with $N$ training points and a substantially larger set of training data being required as one increases the dimensionality of the model. The simplest forms of galaxy--halo connections such as halo occupation distributions have five free parameters \citep{Zheng:2004id}, and it is expected that for more complex selections of galaxy samples the number will grow considerably \citep{Guo:2018wwa,Yuan:2018qek,Favole:2019ouk,Zu:2020uqo}. 

In comparison, modern perturbation theory approaches to galaxy clustering operate at the field level via so-called bias expansions,
which encode the response of small-scale galaxy physics (e.g.\ the galaxy--halo connection) to large-scale structure via a series of bias coefficients (see e.g.\ \citealt{Desjacques:2016bnm} for a recent review). A key advantage of bias models is that while their dependence on parameters is simple and analytic, they should describe the statistics of a broad range of galaxy (and halo) samples as long as they are formed by processes that respect the symmetries of the underlying processes of structure and galaxy formation, namely rotational and Galilean invariance and the equivalence principle. Indeed, it was recently shown that the bias expansion can be directly derived by generating all possible dynamical terms and eliminating combinations not allowed by these symmetries \citep{Fujita20}. \par 

The challenges in using bias models come, instead, from the aforementioned limitations of perturbation theory models themselves. Similarly to perturbation theories for the clustering of dark matter, bias models are not expected to hold across all scales. Instead, they are expected to be valid at scales larger than or comparable to the Lagrangian size of haloes. This regime is where one is insensitive to the internal structure of haloes \citep{McDonald_2009, Fujita:2016dne,Lazeyras:2019dcx,Vlah_2016}. It is worth noting, however, that the nonlinear and halo scales are not identical and scale differently with redshift --- at higher redshifts perturbative models may be more limited by the larger Lagrangian radii of (typically more luminous or massive) samples than dynamical nonlinearities, and vice versa at lower redshifts. This distinction is particularly apparent in the Lagrangian basis \citep{Matsubara2008,Vlah_2016}, in which galaxy clustering due to dynamics and biasing are explicitly disentangled. Recently \citet{modichenwhite19} suggested a way to combine the generality of bias expansion-based models with $N$-body simulations in a manner that is particularly suited for emulation, particularly in the regime where dynamics become nonlinear on scales larger than the halo scales of interest. Since higher-order Lagrangian biases have been found in simulations to be small for low and intermediate mass haloes \citep{Abidi:2018eyd,Lazeyras18}, this scheme keeps the dynamical nonlinearities from $N$-body simulations to all orders while including Lagrangian bias only up to second order. \par

In the remainder of this work we concern ourselves with the construction of an emulator for the halo--halo and halo--matter correlations with analytic dependence on bias parameters, extending the method presented in \citet{modichenwhite19} to a generic cosmological parameter dependence which can then be readily used for cosmological clustering analyses. The  structure is as follows: in section \ref{sec:methods} we briefly review the Lagrangian description of galaxy bias. In section \ref{sec:componentspectra} we describe the hybrid technique which combines displacements obtained from $N$-body simulations with Lagrangian bias. Section \ref{sec:sims} describes the \texttt{Aemulus} suite of simulations \citep{DeRose:2018xdj}, which we use to build the training data for the emulator. The measurements of the `basis spectra' of the hybrid Lagrangian bias model, and their emulation, are outlined in section \ref{sec:emu}. Section \ref{sec:results} concerns itself with assessing the performance of the emulator. Specifically, sub-section \ref{subsec:errors} addresses the scale and redshift-dependent error for each of the ten basis functions that span the model. Subsection \ref{subsec:modelfits} assesses how well the model describes the statistics of complicated galaxy samples, including those possessing concentration and spin secondary biases, as well as the effect of baryons at small scales. Our final test, subsection \ref{subsec:inferencetests}, pits the emulator against a series of increasingly complex simulated likelihood analyses, in order to assess potential biases in inferred cosmological parameters using our emulator and their origin. 

\section{Lagrangian bias expansion}
\label{sec:methods}

In the Lagrangian approach to bias formulated in \citet{Matsubara2008}, the observed clustering of galaxies is obtained through first weighting fluid elements by a local functional $F[\delta(\textbf{q})]$ at their initial (Lagrangian) positions $\textbf{q}$ and then advecting these weights to their observed positions via fluid trajectories $\textbf{x} = \textbf{q} + \mathbf{\Psi}$, where $\mathbf{\Psi}(\textbf{q},t)$ is the Lagrangian displacement. As discussed in the introduction, the bias functional $F$ is obtained by summing up all scalar terms allowed by Galilean invariance and the equivalence principle up to a given order in the initial conditions; up to quadratic order we have \citep{Vlah_2016}
\begin{align}
\label{eqn:lagbias}
F(\boldsymbol{q}) \approx\, 1 + &b_1 \delta_L (\boldsymbol{q}) + \frac{b_2}{2!} (\delta_L^2(\boldsymbol{q}) - \langle \delta_L^2 \rangle )\,+\\
\nonumber & b_{s^2} (s_L^2(\boldsymbol{q}) - \langle s_L^2 \rangle)+ \,b_{\nabla^2}\nabla^2 \delta_L(\boldsymbol{q}) + \, \epsilon(\boldsymbol{q}),
\end{align}
where $s^2 = s_{ij} s_{ij}$ is the tidal shear tensor. The bias expansion is local above the halo scale and the initial fields in the above functional are to be interpreted as smoothed; any `nonlocal' effects as we approach this scale, as well as dependences on smoothing, are parametrized to lowest order by the derivative bias $b_{\nabla^2}$. Modes below the halo scale, uncorrelated with the large scales of interest, are represented by the stochastic noise $\epsilon$.

From the weighting $F(\textbf{q})$, the observed clustering is given via number conservation to be 
\begin{equation}1 + \delta_\alpha(\textbf{x},z) = \int d^3 q\, \delta^D (\textbf{x} - \textbf{q} - \mathbf{\Psi}(\textbf{q},z)) F(\boldsymbol{q}),
\end{equation}
where the Lagrangian displacement $\mathbf{\Psi}$ denotes the movement of the fluid element relative to its initial position. At any given order, the Lagrangian galaxy overdensity above can be mapped onto e.g.\ the Eulerian basis of \citet{McDonald_2009} by Taylor expanding $\mathbf{\Psi}$. However, keeping the nonlinear mapping in the integral above will generate a tower of Eulerian bias parameters even if only a few of the Lagrangian bias parameters are nonzero \citep[see e.g.][]{Abidi:2018eyd}.  We will treat the bias values, $b_\alpha$, as free parameters.  Ab initio predictions of the $b_\alpha$ for general tracer populations is a harder problem, and a current active area of research.

\section{Lagrangian bias and simulations}
\label{sec:componentspectra}

Recently, it has been proposed that one can combine the fully resolved dark matter dynamics of an $N$-body simulation with the analytic perturbative bias techniques we outlined in the previous section \citep{modichenwhite19}. The use of dynamics from an $N$-body simulation means this hybrid model circumvents the need for perturbative calculations related to the equations of motion of the dark matter fluid itself.  Additionally, $N$-body simulations are relatively inexpensive (compared to hydrodynamical simulations) and well-controlled, well-defined limits for observables exist so that convergence of measured quantities can be assessed systematically \citep[e.g.][]{Power_2016,Mansfield:2020lyp,Joyce:2020qxv}. As such, this hybrid model combines two techniques with solid theoretical foundations, ensuring robustness of its predictions. We will briefly describe the technique and how one implements it below, but refer the reader to \citet{modichenwhite19} for a more complete discussion.

When creating initial conditions of an $N$-body simulation, one starts from a noiseless linear cosmological density field, $\delta_L (\boldsymbol{x})$. Traditionally, this density is only used to sample initial displacements which impart a cosmological signal on a set of \emph{pre-}initial conditions. First-order displacements using the Zeldovich approximation,
\begin{equation}
    \Psi(\boldsymbol{q}) = \int \frac{d^3 k}{(2\pi)^3} e^{i \boldsymbol{k} \cdot \boldsymbol{q}} \frac{i \boldsymbol{k}}{k^2} \delta_L (\boldsymbol{k}),
\end{equation}
result in so-called 1LPT initial conditions. However, higher order initial conditions \citep{Crocce:2006ve,Garrison:2016vvp,Michaux:2020yis} are now ubiquitous in modern simulations.\par
\begin{figure*}
    \centering
    \includegraphics[width=0.9\textwidth]{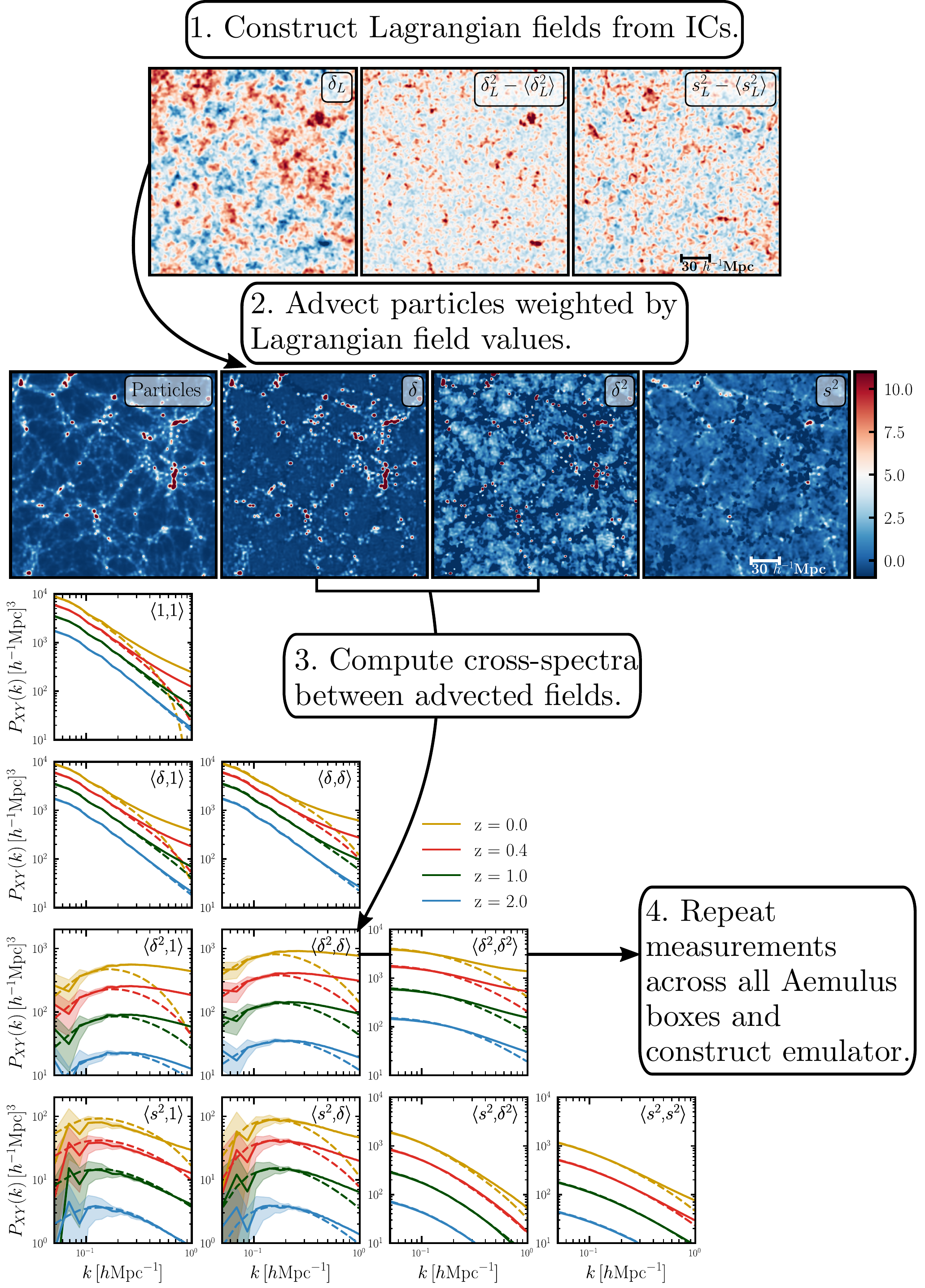}
    \caption{Visualization of the methodology implemented in this paper, from the advection process to the measurements of the basis spectra. Our emulation scheme approximates the cosmology and redshift dependence of each spectrum in the ten panels in the lower part of the figure. The top panel has each Lagrangian field scaled to have equal variance, in order to highlight the qualitative differences between the fields. The middle panel shows the bias weighted-fields that result from the advection process. Different weights highlight qualitatively different aspect of the matter density. The cross-spectra of these fields give the spectra shown in the lower panel.
    \label{fig:fieldlevel}}
\end{figure*}

The noiseless initial density field can also be used to construct the different component fields of the Lagrangian bias expansion of the initial conditions:
\begin{equation}
    {O}_L \supset {1, \delta_L, \delta_L^2, s_L^2, \nabla^2 \delta_L, \cdots},
\end{equation} 
where the subscript $L$ indicates these are the Lagrangian fields. Advecting $N$-body particles weighted by $\mathcal{O}_L$ to a specific snapshot results in bias-weighted fields,
\begin{equation}
    \delta_{\mathcal{O}_L}(\textbf{x}) \equiv \int d^3\textbf{q}\ \mathcal{O}_L(\textbf{q})\ \delta_{D}(\textbf{x} - \textbf{q} - \Psi(\textbf{q})), 
\end{equation}
which trace the non-linear dark matter distribution. In Fig.\ \ref{fig:fieldlevel} (middle panel) we show an example of the different bias-weighted fields produced by this procedure. These fields are similar to the `Eulerian-shifted' operator basis of \citet{Schmittfull_2019}. A notable difference is that in our case the displacements are fully resummed, while the Eulerian-shifted basis of \citet{Schmittfull_2019} only resums the Zeldovich displacement (1LPT). Higher order displacements ($n$LPT) are Taylor-expanded up to third order as part of their bias expansion. The difference is because our aim in this paper is to attempt to model scales beyond the reach of standard one-loop perturbation theory, whereas the goal of \citet{Schmittfull_2019} was to validate one-loop perturbation theory at the field level (see also \citealt{2018PhRvD..98j3532T}). \par 
The power spectrum of any combination of tracers can then generically be written as ($X,\,Y \equiv {\delta_{\mathcal{O}_L}}$)
\begin{equation}
    P^{ab}(k) = \sum_{X,Y} b_{X}^a b^b_{Y}P_{XY}(k) + P_{SN},
\label{eqn:pab}
\end{equation}
where $P_{XY}$ is the cross-power spectrum at a fixed cosmology between the different fields at a given redshift.  For example, the unweighted spectrum, $P_{11}$, is the non-linear matter power spectrum. \par 
This Lagrangian bias model can handle cross-correlations of arbitrary tracers. However, we also note that given a set of bias parameters for a single tracer sample $\alpha$, $\{b_X^\alpha,\, X \in \mathcal{O}_L \}$, one can also self-consistently predict the tracer--matter cross-correlation by taking the second sample to have $b_Y^{m}=0$ except for $Y=1$.  In this case there are only $P_{X1}$ terms. The tracer--matter cross-correlation is the primary cosmic contribution to the signal of galaxy--galaxy lensing, one of the key cosmological observables of current and upcoming galaxy surveys \citep{ Prat_2018, Yoo_2006, Wibking:2019zuc, Mandelbaum:2017jpr}. The tracer--matter cross-correlation is also the primary contribution to the cross-correlation between galaxy positions and lensing of the cosmic microwave background (CMB), one of the most powerful and complementary statistics that is measured between galaxy and CMB surveys \citep{Bianchini_2015, 2016MNRAS.460.4098P, 2017MNRAS.469.4630D, 2018MNRAS.481.1133P, Omori_2019,Singh_2019, 2020JCAP...05..047K}. For notational convenience, throughout the remainder of this paper we will refer to the tracer--tracer correlation as $P^{hh} (k)$ and the tracer--matter correlation as $P^{hm} (k)$. \par 
This hybrid approach of combining $N$-body simulations with Lagrangian bias can fit the power spectrum of tracers to significantly smaller scales than standard Lagrangian perturbation theory \citep{modichenwhite19}. While the dependence on the Lagrangian bias parameters $b_X$ is analytic in this model, one still requires an $N$-body simulation to measure the basis spectra. An $N$-body simulation at a given point of cosmological parameter space then provides a measurement of the basis spectra at that point. With $N$-body simulations that sufficiently sample parameter space one can estimate the cosmological dependence of these basis functions across the entire space. This is precisely the goal of this work.

\section{The \texttt{Aemulus} simulations}
\label{sec:sims}
In order to properly emulate the cosmology dependence of the basis spectra $P_{XY}(k)$, the underlying suite of $N$-body simulations used for measurements of observables must be constructed carefully. \par 
The \texttt{Aemulus} suite of $N$-body simulations \citep{DeRose:2018xdj} has been purpose-built for precise emulation of cosmological observables measured in galaxy surveys. The suite is composed of a set of 75 simulations that span 47 points in the $w$CDM parameter space allowed by a combination of modern CMB, BAO and type Ia supernova experiments. \par 
Each \texttt{Aemulus} box has a size $L_{\rm box} = 1050\,h^{-1}$Mpc with $N=1400^3$ particles, corresponding to a mass resolution of $3.51 \times 10^{10} \left ( \frac{\Omega_m}{0.3} \right)h^{-1} M_\odot$. The \texttt{Aemulus} simulations have undergone rigorous convergence and validation tests for several observables. There are 10 particle snapshots ranging from $0 < z < 3$, allowing for measurements the redshift-dependence of the non-linear basis spectra. \par 
\texttt{Aemulus}' halo mass function emulator has sufficient accuracy to remain valid, for the defined cosmological parameter space, through the Rubin Observatory's Y1 LSST survey \citep{McClintock:2018uyf}, while the galaxy correlation function can predict the clustering of massive galaxy samples, such as those observed by DESI, to within 1 per cent down to scales of $r\approx 1\,h^{-1}$Mpc \citep{Zhai:2018plk}. \par 
Thus, \texttt{Aemulus} represents an appropriate setting to construct an emulator for the Lagrangian bias basis spectra described in section \ref{sec:componentspectra}. The only missing component is that the initial conditions code used in \texttt{Aemulus}, \texttt{2LPTIC} \citep{2012ascl.soft01005C}, does not output the noiseless linear density fields. We patched the code to read out this field and re-generated the initial conditions. \par
\begin{figure*}
    \centering
    \includegraphics[width=\textwidth]{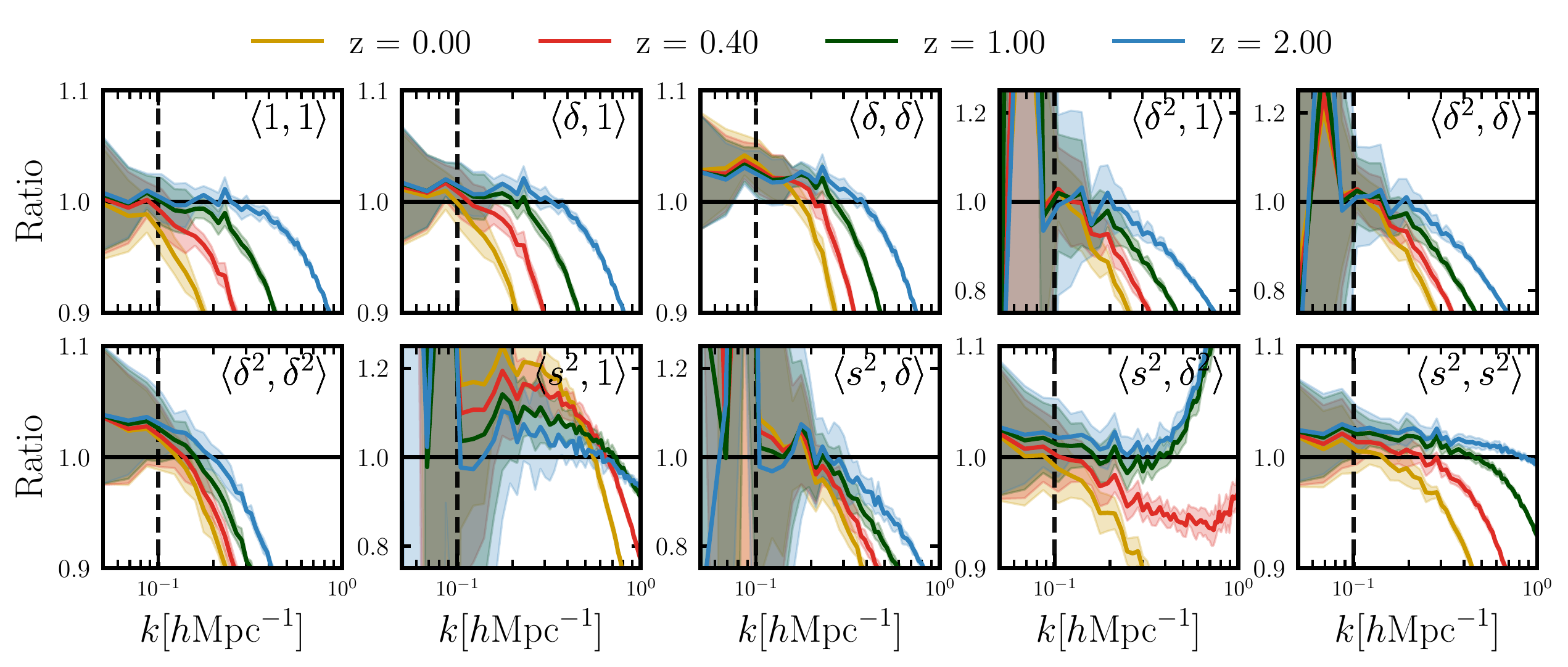}
    \caption{Ratio of the measured basis spectra compared to LPT predictions for one of the cosmologies in the Aemulus test set. The mean of the five independent boxes in the test set is shown, and the shaded band represents one standard deviation as inferred from the boxes. The dashed vertical line at $k = 0.1 h {\rm Mpc}^{-1}$ shows the point where we revert to predictions of LPT. As discussed in the text, we find some small multiplicative differences at large scales for most basis spectra, that are larger for the basis spectra built from higher powers of the density field. This is most likely due to discrepancies in growth factors obtained between linear theory and $N$-body simulations.
    }
    \label{fig:lptratio}
\end{figure*}
\section{Emulating the basis spectra} 
\label{sec:emu}
\subsection{Measuring basis spectra}

We now describe in detail our implementation of the hybrid Lagrangian biasing scheme described in Section \ref{sec:componentspectra}. Schematically, the process of obtaining measurements of the basis spectra from an $N$-body box can be broken down into four steps: 
\begin{enumerate}
    \item Compute the Lagrangian bias fields: given the noiseless density field $\delta_L$ one constructs the other weight fields $\mathcal{O}_L$ by applying the appropriate transformations.
    \item Advect particles to a given snapshot: every particle ID can be associated with a grid cell $\{i,j,k\}$ in the fields $\mathcal{O}_L$. Every particle in a snapshot receives a weight $\left (\frac{D(z)}{D(z_0)} \right)^n \times \mathcal{O}_{L}[i,j,k]$, where $\left (\frac{D(z)}{D(z_0)} \right)$ is the ratio of growth factors between the snapshot and initial conditions, and $n$ is the number of powers in the linear density field that make up $\mathcal{O}_L$.
    \item Paint the weighted particles to a grid, to form the late-time bias fields.
    \item Measure the basis spectra: the painted bias fields are cross-correlated with each other to measure the basis  spectra $P_{XY}$ for that given cosmology and redshift.
\end{enumerate}

This procedure imposes some additional storage requirements. While a particle catalog normally has seven entries for every particle, $(ID, \, \boldsymbol{x}, \boldsymbol{v})$, each bias field weight will add an additional entry. Naively saving component weights at every snapshot will lead to a 57 per cent increase in catalog size. However, the time evolution of the weights is determined entirely by the linear growth function and can be determined on the fly. Thus, the fractional increase in catalog size will only be of order $\sim (1/7)(N_b/N_z)$, where $N_b$ is the number of bias-weighted fields computed and $N_z$ is the number of snapshots used.  For the second order basis of $\mathcal{O} = \{1, \delta_L, \delta_L^2, s_L^2, \nabla^2 \delta_L\}$ this represents a fractional increase in catalog size of 6 per cent.
Even if the weights are not stored, all of the steps outlined above can be carried out on the fly when needed.\par 
In Fig.~\ref{fig:fieldlevel} we show a comparison between the predictions of one-loop Lagrangian perturbation theory and the basis spectra averaged across five \texttt{Aemulus} boxes with the same cosmology, from the test suite. For all basis spectra we recover the LPT result at large scales to within a few per cent. While one would expect the agreement at large scales to be exact, it is well known that $N$-body simulations struggle to correctly recover linear growth at large scales \citep{Heitmann:2008eq,Schneider_2016,Garrison:2016vvp} due to transients from the grid that particles are initialized on, and the discrete nature of the kick-drift-kick operators used in time-stepping. This discrepancy is also present in \texttt{Aemulus}, as can be seen in fig.~13 of \cite{DeRose:2018xdj}. The \texttt{Aemulus} simulations have a 1 per cent mismatch in growth at large scales, which is redshift independent at the largest scales. Differences in growth between linear theory and the simulations would then be amplified for the basis spectra built from multiple fields. In Appendix \ref{appendix:lptrecover} we explore the $k\to0$ differences between LPT and our emulator, present prescriptions for enforcing consistency and discuss the small impact they have on parameter inference. \par 
At small scales, we see that non-linear structure formation imbues significant differences between LPT and the simulations. At the highest redshift shown, $z=2$, the agreement for the three spectra that dominate the signal ($\langle 1,1 \rangle, \, \langle 1, \delta \rangle, \mathrm{ and } \langle \delta, \delta \rangle$) is close throughout all scales probed in our simulation. Thus, for the scales under consideration, we find no need to extend the emulator to $z > 2$.\par 

Above $z=1$, the \texttt{Aemulus} simulations only have snapshots at $z=2$ and $z=3$, and thus any attempt to emulate redshift evolution between these snapshots is too poorly sampled for the emulator to achieve our desired performance. For $z \geq 2$ the emulator reverts to predictions from \texttt{velocileptors} \citep{Chen:2020fxs}, a public code to predict LPT power spectra and correlation functions to one loop order. This agrees quite well with most basis spectra given Fig.~\ref{fig:fieldlevel}. When reverting to LPT at $z>2$, our implementation includes an additional free parameter. This parameter corresponds to the $k^2$ counterterm for matter that takes into account the effects of small-scale physics not captured by perturbation theory \citep{Vlah15}. We note that there are no specific impediments to measuring basis spectra, or the emulation scheme adopted, at higher redshifts. Given simulations that are sufficiently well sampled in time, out to the furthest bin one wishes to include, the techniques described here should apply.\par 
The LPT predictions shown in Fig.~\ref{fig:fieldlevel} are a limit of a more complete theory that includes redshift-space distortions \citep{chen2020redshiftspace,Chen:2020fxs} . The agreement between $N$-body simulations and this subset of LPT at large scales implies the bias parameters in the full theory and our hybrid model are equivalent; a set of bias parameters obtained from fitting the emulator to a sample can then be used in tandem with RSD measurements analysed purely with perturbation theory at a slightly more restrictive $k_{\mathrm{max}}$.  Since the RSD measurements are done in 3D, rather than projection, one can achieve small measurement errors at more restrictive $k_{\rm max}$ making this combination an efficient one, e.g.\ for testing general relativity \citep{Alam:2016qcl,Zhang:2020vru}. \par 
We note that we omit results for the basis spectra $\langle X, \nabla^2 \delta \rangle$. The initial weight field $\nabla^2 \delta_L $ has a large amount of power at very small scales, making its Fourier transform unwieldy due to the presence of an explicit smoothing scale of $k\sim L_{\rm grid}^{-1}$. As a result, we find the basis spectra as measured through the advection procedure have a cosmology-dependent amplitude mismatch when compared to LPT predictions at large scales. Therefore we adopt the approximation $\langle X, \nabla^2 \delta \rangle \approx -k^2 \langle X, 1 \rangle$ in the actual emulation scheme. Since these higher derivative bias contributions most closely correspond to the effects of baryonic physics and finite-size effects for haloes, we check that the approximation performs similarly in Section \ref{subsec:modelfits}. Specifically, in Fig.~\ref{fig:baryons} we explicitly show the differences between the measured $P_{1\nabla^2}$ and the approximation employed.
We also note the approximation lowers the complexity of the emulation scheme, reducing the full set of basis functions at second order to be emulated from  15 to 10.

\subsection{Principal components of non-linear spectra}

Once the basis spectra have been measured across all boxes, the emulator is built by adopting a suitable interpolation scheme between the different spectra. While other emulators using the \texttt{Aemulus} simulations have been constructed using Gaussian processes (GPs), we adopt a different approach here, similar to that used in the Euclid emulator \citep{Knabenhans:2018cng}, using a combination of principal component analysis and polynomial chaos expansions (PCE) \citep{10.5555/1893088}.

We prefer PCE to GP emulation for a few practical reasons. GPs are more difficult to train, requiring explicit choices for kernels and tuning of real valued hyper--parameters. Additionally, the run time for evaluating a trained GP scales with the amount of data used for training, while the run-time of a PCE model evaluation scales only with the order of the PCE. Furthermore, the polynomial nature of PCEs means that they have fast, analytic gradients, making them easy to integrate with sampling techniques such as Hamiltonian Monte Carlo \citep{HMC}, although we have not done so in this work. GPs may still be preferred when the model being emulated is highly complex, but, as we show in the following sections, we are able to attain a nearly optimal emulator performance with the simpler and faster PCE scheme.

To begin, we compute one-loop LPT predictions for each basis spectrum at every cosmology and redshift in the \texttt{Aemulus} training design, which we will refer to as $P_{\rm XY}^{\rm LPT}(k,\mathbf{\Omega})$, where $\mathbf{\Omega}$ denotes the cosmology and redshift in question. To do this we make use of the \textsc{velocileptors} code \citep{Chen:2020fxs}. 

We then compute the ratio between the LPT predictions and the measured basis spectra, $P_{XY}^{\rm NL}(k,\mathbf{\Omega})$, from each snapshot. These ratios are thus consistent with unity at small wavenumbers, and while they deviate significantly from unity at high $k$ they have significantly less dynamic range than the basis spectra. In order to de-noise these ratios, we apply a Savitsky--Golay \citep{doi:10.1021/ac60214a047} filter of order three using an 11-point window in $k$. Doing so dramatically reduces the amount of noise in the spectra, and is a simple alternative to reduce noise at high $k$, where techniques such as fixed amplitude, paired phase simulations do little to reduce variance \citep{Angulo:2016hjd, Villaescusa-Navarro:2018bpd, Chuang:2018ega}. As a final preprocessing step, we also take the base-10 logarithm of these smoothed ratios in order to further decrease the dynamic range. This yields the quantity that we emulate, which we call $\Gamma^{XY}(k, \mathbf{\Omega})$,
\begin{figure}
    \centering
    \includegraphics[width=\columnwidth]{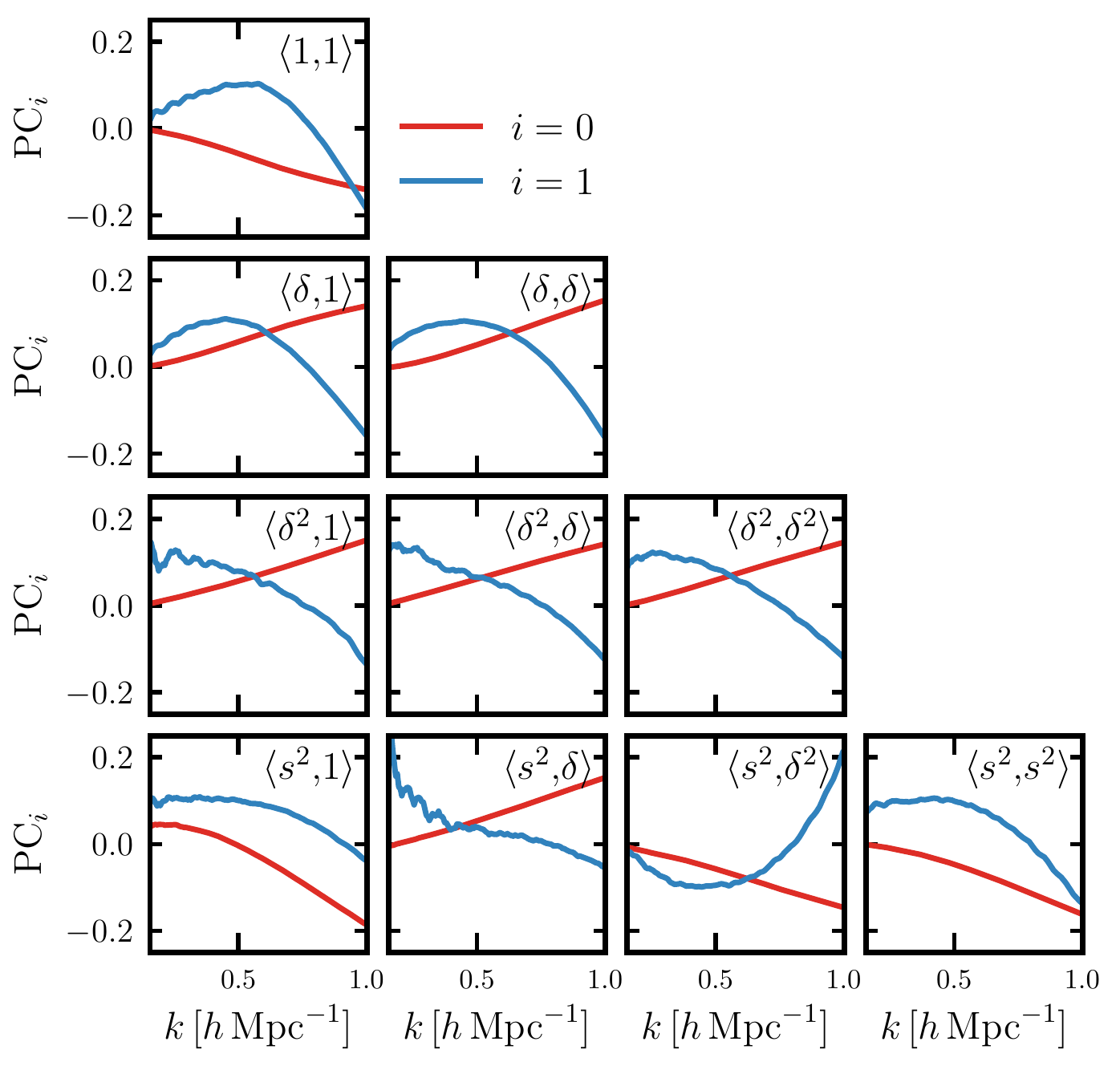}
    \caption{The first two principal components of the log-ratios between $N$-body and LPT spectra, $\Gamma^{XY}$, for each basis spectrum. The principal components are very smooth compared to the raw basis spectrum measurements from the simulations. Two principal components are sufficient to explain greater than 99 per cent of the variance in all spectra as a function of redshift and cosmology.}
    \label{fig:pcs}
\end{figure}
\begin{equation} 
\Gamma^{XY} (k, \mathbf{\Omega}) \equiv \log_{10} \left ( \frac{P_{XY}^{\rm NL}(k,\mathbf{\Omega})}{P_{XY}^{\rm LPT}(k, \mathbf{\Omega})} \right)
\end{equation}
After these pre-processing steps, we proceed by constructing a principal component basis for these spectra. At this point we restrict ourselves to $0.1 < k < 1$ and $0 < z < 2$, however we note that in principle there are no issues extending to broader scales and redshifts if simulations allow for it.\par 
Let $\mathbf{X}_{XY}$ be the $N \times M$ array containing $\Gamma^{XY}$, where $N=N_{\rm cosmo} \times N_{z}$, $N_{\rm cosmo}$ is the number of cosmologies in our training set, $N_z$ is the number of redshift outputs per cosmology in our training set and $M$ is the number of $k$ values under consideration. Then a basis of principal components can be constructed by computing the eigenvectors of the covariance matrix of $\mathbf{X}_{XY}$:

\begin{align}
    \mathbf{C}_{XY} &= \mathbf{X}_{XY}^{\rm T} \mathbf{X}_{XY},\\
    &= \mathbf{W}_{XY}\mathbf{\Lambda}_{XY} \mathbf{W}_{XY}^{\rm T},\nonumber
\end{align}
where the rows of $\mathbf{W}_{XY}$ are the eigenvectors, i.e., the principal components, in question and $\mathbf{\Lambda}_{XY}$ is a diagonal matrix of the eigenvalues, which are equal to the variance of the data described by each eigenvector. In all cases, greater than 99 per cent of the variance in each basis spectrum is described by the first two principal components, shown in Figure \ref{fig:pcs}. We thus disregard all other principal components for the duration of this work. Given the results discussed in Section \ref{sec:results}, we deem this to be sufficient. Having computed the principal components, we then determine the projection of them onto each measured $\Gamma^{XY}$ via:

\begin{align}
    \mathbf{A}_{XY} = \mathbf{X}_{XY}\mathbf{W}_{XY},
\end{align}
where $\mathbf{A}_{XY}$ is an $N\times2$ matrix containing the principle component coefficients $\alpha^{XY}_{i}(\mathbf{\Omega})$ for each cosmology and redshift in our training set. It is the dependence of these coefficients on cosmology and redshift that we build a surrogate model for using polynomial chaos expansions \citep{10.2307/2371268}. \par 
\subsection{Emulating cosmology dependence with polynomial chaos}
With our principal components in hand, every point in cosmological parameter space sampled by the training set has coefficients for the approximation
\begin{align}
    \Gamma^{XY}(k, \mathbf{\Omega}) \approx \sum_{i} \alpha_i^{XY} (\mathbf{\Omega}) \mathrm{PC}_i^{XY} (k).
\end{align}
The problem of emulating the cosmology dependence of the $\Gamma^{XY}$ functions is now reduced to that of figuring out the cosmology dependence of the PC coefficients $\alpha_i (\mathbf{\Omega})$. A polynomial chaos expansion (PCE) (of order $N$) of this dependence is the decomposition of the $\alpha_i$ onto a basis of products of orthogonal polynomials $\Phi_{\mathbf{i}} (\mathbf{\Omega})$ organized by a multi-index $\mathbf{i}$ \citep{10.5555/1893088}:

\begin{align}
    \alpha (\mathbf{\Omega}) =  \sum_{|\mathbf{i}| \leq N} c_\mathbf{i} \Phi_\mathbf{i} (\mathbf{\Omega}).
\end{align}
Each component of the multi-index $\mathbf{i} = (i_1,\cdots,i_d )$, denotes the order of the polynomial for that cosmological parameter, e.g., 
\begin{align}
    \Phi_\mathbf{i} (\mathbf{\Omega}) =  \phi_{i_1} (\Omega_1) \cdots \phi_{i_d} (\Omega_d),
\end{align}
and so $\phi_{i_d} (\Omega_d)$ is a univariate orthogonal polynomial of order $i_d$. \par 
While this is in principle a decomposition into a combinatorially large space of coefficients $c_\mathbf{i}$, it is known to be a sparse representation \citep{2008CRMec.336..518B,2011JCoPh.230.2345B}, and there exist many algorithms (and numerical libraries) optimized to perform regression over this space and obtain values for the coefficients. We use the package \texttt{Chaospy} \citep{FEINBERG201546,doi:10.1137/15M1020447} to perform the decomposition and subsequent regression. Note that since the parameter dependence of the principal components is given by a combination of polynomials, our model in principle has an analytic dependence on cosmology, redshift, and bias.
Since the coefficients are determined via regression, a PCE emulator does not recover the input data exactly. However, the tests conducted in section \ref{sec:results} indicate that this drawback is not an issue.\par
In total, the hyperparameters in the model are:
\begin{enumerate} 
\item The number of principal components used, $N_{\mathrm{PC}}$.
\item The maximum order of the multi-index $|\mathbf{i}|$. In practice we separately optimize over the maximum polynomial order of each individual parameter $i_{d}$, with $i_d \leq 4$.
\end{enumerate}
As mentioned previously, we restrict ourselves to $N_{\mathrm{PC}} = 2$, as this is sufficient to capture over 99 per cent of the variance in each basis spectrum. To optimize over the polynomial orders $i_d$, we run a simple grid search across the aforementioned values for the seven $w$CDM parameters $\mathbf{\Omega} = (\Omega_b h^2,\Omega_c h^2, \sigma_8, H_0, n_s, N_{\mathrm{eff}}, w)$ and evaluate our results on the \texttt{Aemulus} test suite. We select the set of orders that minimizes global error across all test boxes and snapshots. We describe the tests of this optimized emulator below.
\begin{figure}
    \centering
    \includegraphics[width=\columnwidth]{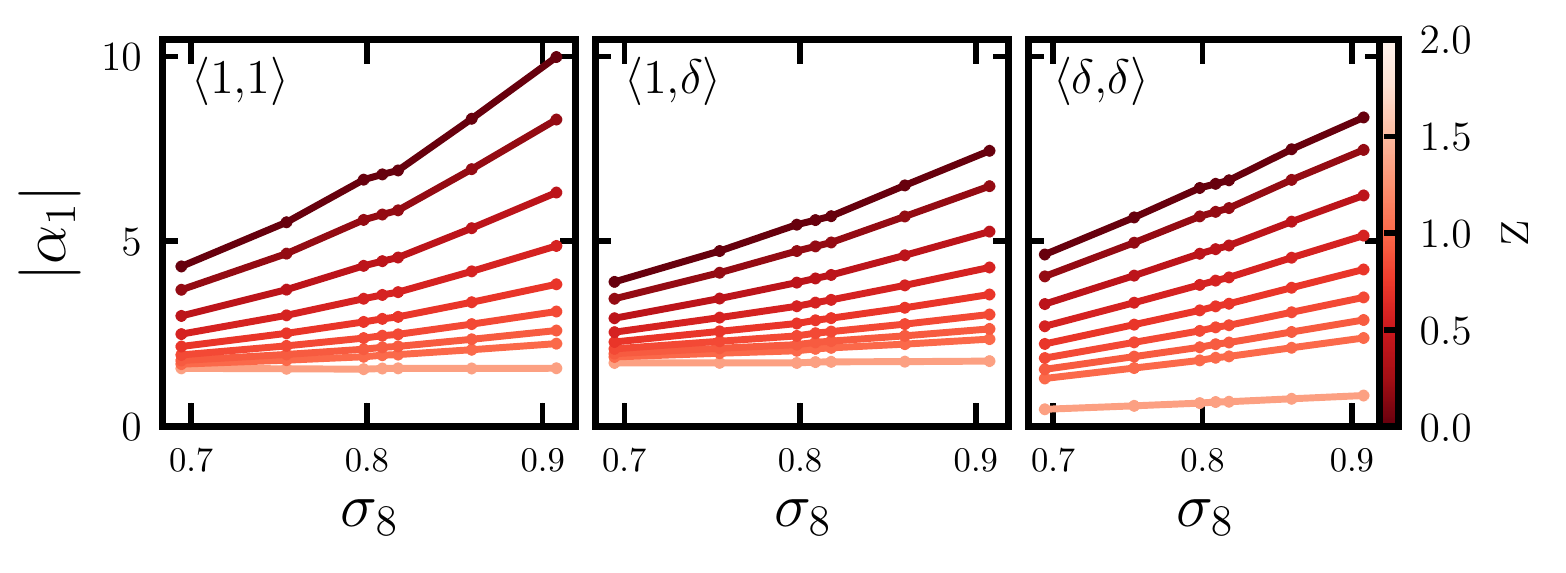}
    \caption{Coefficients of $\mathrm{PC}_1^{XY} (k)$ for the first three basis spectra as a function of $\sigma_8$, colored by redshift. The coefficients vary smoothly for all redshifts as $\sigma_8$ is varied. It is the dependence of these coefficients that we emulate via PCE as a function of cosmology and redshift. The panels look similar for the remaining basis spectra.}
    \label{fig:pc_coeffs}
\end{figure}
\begin{figure*}
    \centering
    \includegraphics[width=\textwidth]{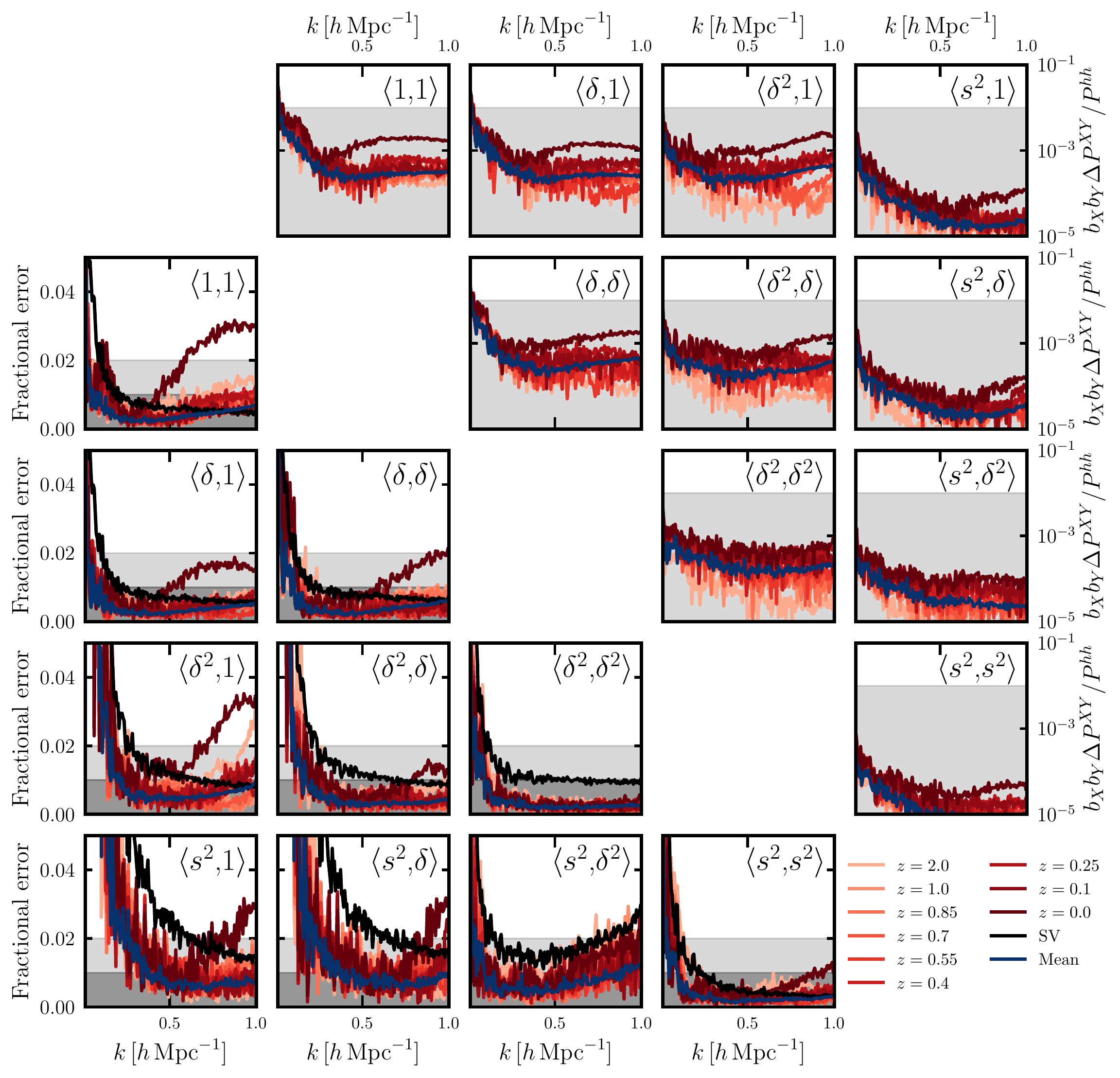}
    \caption{Emulation residuals for basis spectra. 
    \emph{Lower left triangle:} the fractional error obtained for each basis spectrum when compared to the measurements averaged from each set of boxes in the test suite. \emph{Upper right triangle:} the relative size of the emulator residuals compared to the total halo--halo spectrum measured for a fiducial halo sample. In each panel, the dark blue curves are the mean residuals across all redshifts and test boxes, the red curves report the median residual error across the test suite as a function of redshift, and the black curves report the expected sample variance at the volume of an \texttt{Aemulus} training box. }
    \label{fig:residuals}
\end{figure*}
\section{Results}
\label{sec:results}
\subsection{Analysis of emulator residuals}
\label{subsec:errors}
A crucial step in producing viable emulators of cosmological observables is characterizing the accuracy of the emulation scheme. We use the \texttt{Aemulus} set of test boxes to assess the performance of the scheme described in the previous section. The test boxes span seven points in cosmological parameter space, each with five independent realizations of that cosmology. We use the average of five basis spectra at each test cosmology as reference quantities to understand the errors induced in the emulation procedure as a function of scale, across parameter space.\par 
We report the accuracy of our optimized PCE emulator for the basis spectra over the range $0.1 \leq k \leq 1.0\,h\,{\rm Mpc}^{-1}$ in the lower left panel of Fig.~\ref{fig:residuals}. Across most redshift bins in the test suite and for most basis spectra we achieve better than 1 per cent accuracy in the test set. At $z=0$ we observe worse performance, however this can be attributed to numerical difficulties in computing the LPT spectra at $z=0$ at small scales, as can be seen in Fig.~\ref{fig:fieldlevel}. 
As there is little cosmological information in the very low redshift universe, we do not consider this to be a significant issue. Indeed, our additional validation tests support that the model has sufficient accuracy to analyse current survey data.\par 
Adopting a fiducial set of bias parameters corresponding to a halo sample of $12.5 \leq \log_{10} \left ( \frac{M}{h^{-1} M_\odot } \right) \leq 13$, we compute the emulator residuals for each basis spectrum relative to the \emph{total} $P^{hh}(k)$. The results are shown in the upper right triangle of Fig.~\ref{fig:residuals}. The individual basis spectrum error rarely exceeds a permille of the total power. This implies that the slightly larger errors for cubic basis spectra shown in Fig.~\ref{fig:residuals} are sub-leading relative to the total signal we expect to model.

\subsection{Fitting assembly bias and baryons}
\label{subsec:modelfits}

Beyond samples of fixed halo mass, the general bias expansion in Eq.~\ref{eqn:lagbias} should also be able to describe the clustering statistics of more complex tracer populations. It is well known that haloes of a fixed mass bin exhibit different clustering properties depending on whether they are sub-selected on certain properties. This effect, originally discovered in the context of assembly history, and generally known as assembly bias or secondary bias, has been observed for selections on concentration, occupation, local environment, spin, and other secondary halo properties \citep{Wechsler:2001cs, Gao:2005ca, Wechsler:2005gb,Dalal:2008zd,Mao:2017aym,Salcedo_2018,Mansfield:2019ter}. \par 
\begin{figure*}
    \centering
    \includegraphics[width=\textwidth]{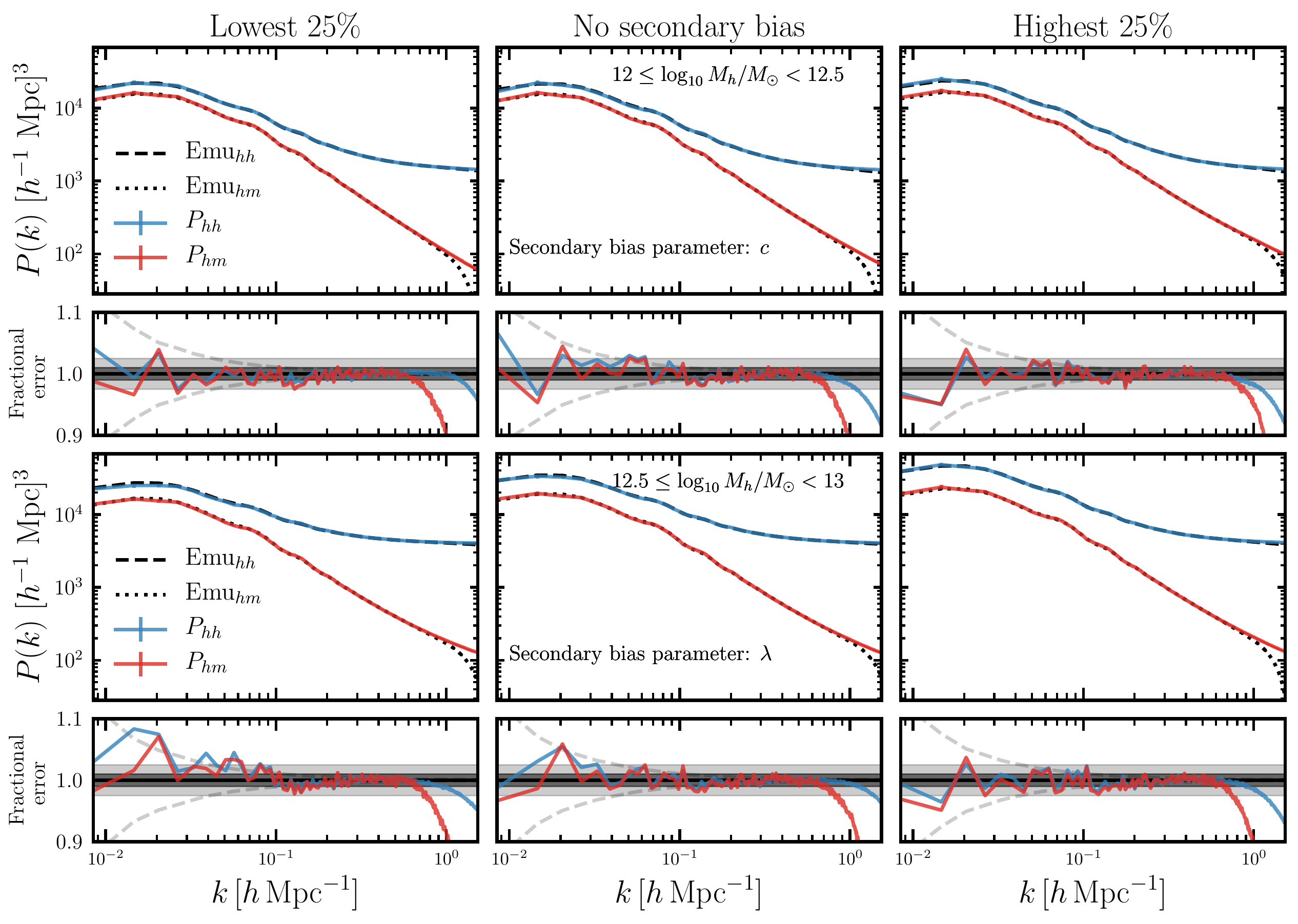}
    \caption{Emulator predictions at fixed cosmology for halo samples exhibiting concentration (top panels) and spin (bottom panels) assembly bias. Central panels show the signal from the halo sample with no selection on a secondary parameter. The left and right panels show samples split on the lowest and highest quartiles of the relevant secondary bias parameter, respectively. Shaded bands show the regions where residuals are within 2 per cent and 1 per cent respectively, while the dashed envelope shows the expected cosmic variance for a sample with $V \approx 5.8 (h^{-1} {\rm Gpc})^3$. The spectra are measured at $z=0.7$ and the fit is performed with the data vector out to $k_{\rm max} = 0.6\, h {\rm Mpc}^{-1}$.}
    \label{fig:secondarybiasfits}
\end{figure*}
As a test of our model, we construct halo catalogs with different amounts of concentration and spin secondary bias, splitting the sample by quartile. The magnitude of the effect varies differently as a function of mass for each secondary bias parameter. Thus, we adopt separate halo mass bins for each parameter, in a regime where we have both reliable estimates of the secondary quantities and know that the secondary bias effect is not drastic, following fig. 4 of \cite{Sato-Polito:2018pvu}. The mass range $12 \leq \log_{10} \left ( \frac{M}{h^{-1} M_\odot } \right) \leq 12.5$ was used to build samples contaminated with concentration bias, and $12.5 \leq \log_{10} \left ( \frac{M}{h^{-1} M_\odot } \right) \leq 13$ for spin bias. We consider the highest and lowest quartile samples in both concentration and spin, as well as a sample with no secondary bias, sub-sampled to the same number density as the samples contaminated with secondary bias. We additionally do not subtract the shot-noise contribution from measured spectra, and opt instead to include it in our covariance matrix as detailed in Eqn.~\ref{eqn:cov}. \par 
Using the emulator for the basis spectra evaluated at the cosmology of these test boxes, we jointly fit the halo--halo and halo--matter spectra $\{P_{hh},P_{hm}\}$ with five parameters: $b_i = \{b_1, b_2, b_{s^2}, b_{\nabla^2}, \bar{n}^{-1}\}$. We minimize the $\chi^2$ between the mean of five simulations assuming a disconnected covariance for the observables as described in Eq.~\ref{eqn:cov}, with $V = 5 \times (1.05\,h^{-1} {\rm Gpc})^3$ each. The resulting fits are shown in Fig.~\ref{fig:secondarybiasfits}. \par 
We fit the spectra to a maximum scale of $k_{\rm max} = 0.6\,h\,{\rm Mpc}^{-1}$. For most panels, we see that the hybrid $N$-body/Lagrangian bias model can jointly describe the clustering and lensing spectra to within 1 per cent down to scales even smaller than employed for the model fit. At large scales, the lowest spin assembly bias bin seems to be systematically higher by at most 10 per cent. Changing the $k_{\rm max}$ of the fit down to $0.2 \, h\,{\rm Mpc}^{-1}$ does not qualitatively alleviate the large-scale discrepancies. We observe similar behavior if the average of the basis spectra from this cosmology are used instead of the emulator, implying this is not an issue of the emulator and could perhaps be attributed to large-scale noise. Another possibility is that a second-order Lagrangian bias model is unable to fully capture the effects of spin secondary bias, but we leave this investigation to future work. \par
\begin{figure}
    \centering
    \includegraphics[width=\columnwidth]{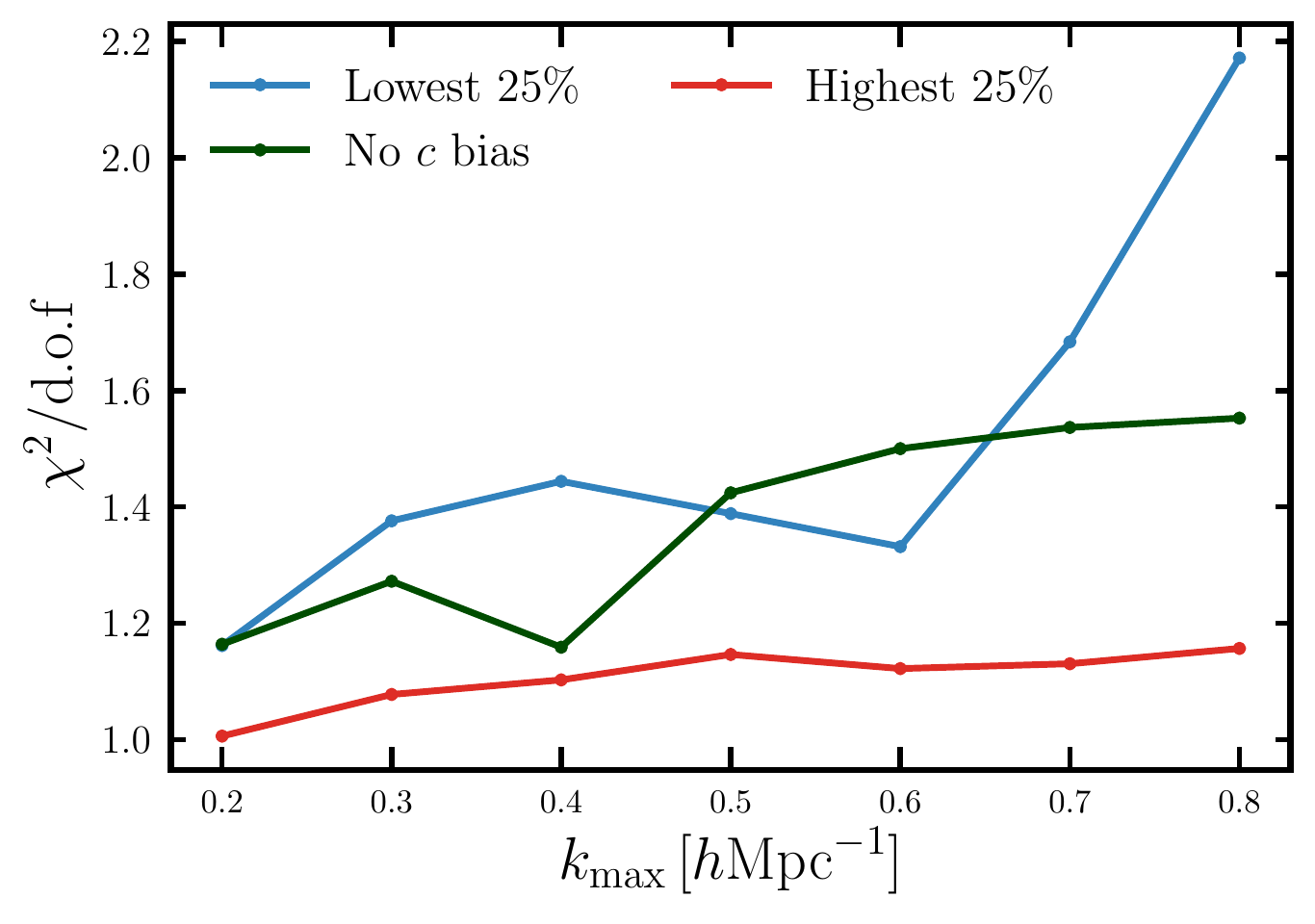}
    \caption{The goodness of fit $\chi^2 / {\rm d.o.f.}$ from increasing $k_{\rm max}$ for the halo sample selected on concentration quartiles, using the emulator as a model. Note the significant degradation of the goodness of fit for the subsample split on the lowest quartile after $k_{\rm max} = 0.6$. }
    \label{fig:chi2plots}
\end{figure}

 
In Fig.~\ref{fig:chi2plots} we show the reduced $\chi^2$ for the fits to the samples split on concentration. We see that the goodness of fit degrades significantly past $k \simeq 0.6  h\,{\rm Mpc}^{-1}$ for some subsamples. The fits to smaller $k_{\rm max}$ have $\chi^2 / {\rm d.o.f.} \lesssim 1.5$. Note that in these tests we use the emulator at a volume that is significantly larger than the boxes it was trained on, and the covariance matrices do not have any contributions due to the emulator uncertainty. If we instead use the mean basis spectra the $\chi^2 / {\rm d.o.f.}$ cross the $\chi^2 / {\rm d.o.f.} \sim 1 $ threshold at $k_{\rm max} \sim 0.6  h\,{\rm Mpc}^{-1}$ and grow significantly afterwards, signalling a potential breakdown of the applicability of this Lagrangian bias model to these samples.\par 
\begin{figure}
    \centering
    \includegraphics[width=\columnwidth]{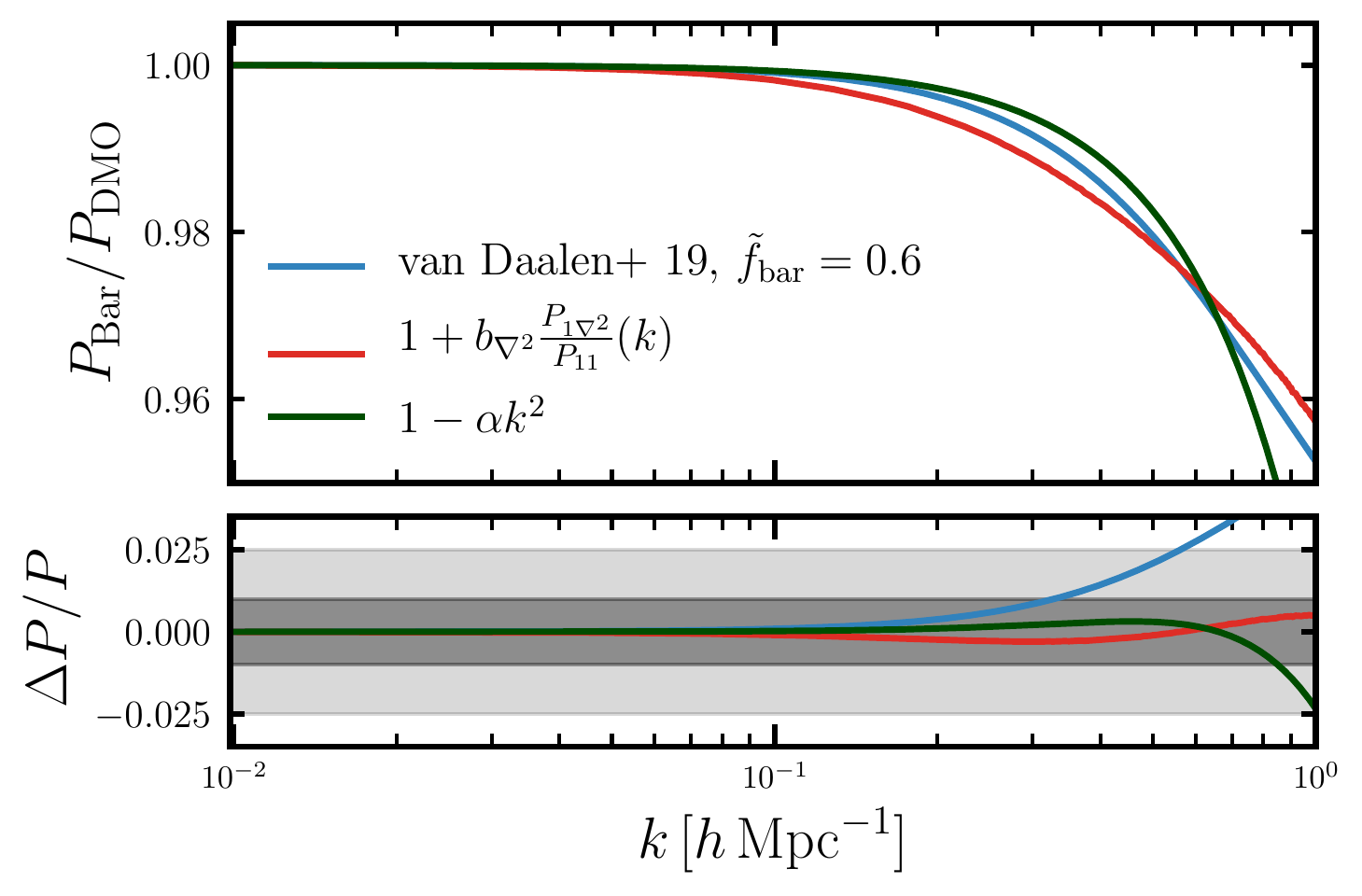}
    \caption{Higher derivative bias terms and their comparison to the baryonic physics fitting function of \citet{vanDaalen:2019pst}. The top panel shows the fitting function, the basis spectrum as measured in the $N$-body simulations and the approximation we employ in the text. In the lower panel we show residuals between the different treatments and the fitting function. The blue curve in the lower panel shows the difference between the unprocessed dark matter power spectrum and the fitting function. The green curve is the approximation to the higher derivative fitting functions that is implemented in our analyses.}
    \label{fig:baryons}
\end{figure}

Baryonic physics is known to impact the statistics of biased tracers at the scales we are considering \citep{White:2004kv,Zhan:2004wq,Chisari:2019tus,vanDaalen:2019pst}. In our model, the $\langle 1, \nabla^2 \delta \rangle$ basis spectrum should have the scale dependence required to capture the first-order impacts of baryons \citep{Lewandowski:2014rca}. In order to test this, we produce mock `baryonified' spectra using the fitting function of \citet{vanDaalen:2019pst}, which is obtained from analysis of a comprehensive suite of hydrodynamic simulations. We compare the fitting function to two parametrizations for the impact of baryons: 
\begin{enumerate}
    \item Including terms that scale as the basis functions $b_{\nabla^2}\langle 1, \nabla^2 \delta \rangle$ and $b_1 b_{\nabla^2} \langle \delta, \nabla^2 \delta \rangle$.
    \item Same as above, but substituting the basis functions with the approximation $\langle X, \nabla^2 \delta \rangle \simeq -k^2 \langle X, 1 \rangle $.
\end{enumerate}
The results of this test are shown in Figure \ref{fig:baryons}. While the baryonic suppression factors presented by the two parametrizations differ, in the bottom panel we see that both capture the effects of baryons to within 1 per cent out to $k \approx 0.8\,h\,\mathrm{Mpc}^{-1}$, whereas not including the contributions leads to errors larger than 1 per cent at $k \approx 0.2\,h\,\mathrm{Mpc}^{-1}$. \par 
Additionally, our framework can simultaneously treat the effects of finite halo size and baryonic physics. As both are captured by the same basis spectra, this corresponds to treating the halo tracer as having one set of $b_\nabla^2$ and the matter tracer in the $P^{hm}$ correlation as having a separate higher derivative coefficient $b'_{\nabla^2}$, while keeping all other bias parameters equal to zero.

\subsection{Recovering input cosmology}
\label{subsec:inferencetests}

In this section we present an increasingly complex series of tests to ensure our emulator can be used for cosmological inference, i.e.,  to demonstrate that it can recover input cosmological parameters in an unbiased way. The general structure of the analyses we run is as follows. \par 
The input data-vectors will be the joint halo--halo and halo--matter power spectra  $\mathbf{d} = \{ P_{hh} (k), P_{hm} (k) \}$. We assume a Gaussian likelihood in the residuals between $\mathbf{d}$ and the emulator prediction at a cosmology $\mathbf{x}(\mathbf{\Omega})$:
\begin{equation}
    \log \mathcal{L}(d | \mathbf{\Omega} ) \propto -(\mathbf{d} - \mathbf{x}(\mathbf{\Omega}) )^T \mathbf{C}^{-1} (\mathbf{d} - \mathbf{x}(\mathbf{\Omega}) ).
\end{equation}
We adopt a baseline covariance matrix that includes only dependence on the two-point functions of the tracer density field, known as the disconnected contribution \citep{Li_2019}. The result is a block-diagonal matrix with format
\begin{equation}
    \mathbf{C}(k, k') \equiv  \frac{2\pi^2\delta_{k, k'}}{k^2 \Delta k V} \times \begin{cases}
    2 P_{hh}^2 (k), &\text{ for } hh \times hh\\
    2 P_{hh}(k) P_{hm}(k),  &\text{ for } hh \times hm\\
    
    \bigg[ P_{hh}(k) P_{mm}(k)   &\text{ for } hm \times hm\\ 
    + P_{hm}^2(k)\bigg],

    \end{cases}
\label{eqn:cov}
\end{equation}
for each sub-block. We use non-linear power spectra and $P_{hh}$ includes the shot-noise contribution. At the smaller scales we probe in the resulting analyses, the purely disconnected approximation is known to fail and off-diagonal (connected) components become increasingly important \citep{Meiksin_1999,Scoccimarro_1999, Cooray_2001, Mohammed_2016, Lacasa_2018}.
The intent of this paper is not to conclusively quantify the information content available at small scales. Rather, we would like to ensure that the emulator is an unbiased model when pushing to such small scales. Therefore, we consider the form of the covariance in Eqn.~\ref{eqn:cov} to be a sufficient baseline to carry out our analyses. We assess its performance in more detail in Appendix \ref{appendix:cov}. 
As we will discuss in more detail in section \ref{subsubsec:haloesample}, the approximation of taking only the disconnected contribution neglects two forms of error: that arising from the connected contribution, and model error from the emulator itself. We discuss the contribution to the covariance from emulator error in Appendix \ref{appendix:cov}, and find that in the regime under which our tests are carried out, its inclusion is important in achieving unbiased constraints. \par 
We sample the posterior distributions of the model parameters via Markov Chain Monte Carlo (MCMC), using \textsc{emcee} \citep{2010CAMCS...5...65G, 2013PASP..125..306F}.  Chains are run with either $N=64$ or $N=128$ walkers across 8000 (4000) steps respectively. We checked that these values ensure converged chains for the simulated likelihood analyses we run; the posteriors are not altered significantly by doubling the length or number of walkers. 
We adopt wide uniform priors on the bias parameters,
\begin{equation}
    b_i \sim U(-5, 5),
\end{equation}
and uniform priors surrounding the boundaries of the \texttt{Aemulus} training suite, specified in Table \ref{table:priors}.
\begin{table}
\centering
 \begin{tabular}{c  c } 
 \hline  \hline 
 Parameter & Range \\
 $\Omega_b h^2  $ & [0.0207 , 0.0237]\\
 $\Omega_c h^2$ & [0.101 , 0.132]\\
 $w_0$ & [-1.399 , -0.566]\\
 $n_s$ & [0.928 , 0.997]\\
 $\sigma_8$ & [0.575 , 0.964]\\
 $H_0$ & [61.69 , 74.77]\\
 $N_{\mathrm{eff}}$ & [2.62 , 4.28]\\ 
 \hline
 \hline
\end{tabular}
\caption{Boundaries of the cosmological parameters of simulations spanned by the \texttt{Aemulus} training suite. These are the values used as flat priors for cosmological parameters.}
\label{table:priors}
\end{table}

\subsubsection{Synthetic Data}
\label{subsubsec:noiselesschain}

As a first test of the emulator, we perform a simulated likelihood analysis on a noiseless data vector drawn from the emulator itself. We fit the basis spectra to a halo sample of mass $12 \leq \log_{10} M_{h}/M_\odot \leq 12.5$ from one of \texttt{Aemulus}' test boxes. The cosmology and best-fitting bias values are used as inputs to the emulator to produce a mock noiseless data-vector. As the data in this test is not a random draw from a distribution, the exact format of the covariance matrix does not matter. However, we use the block-diagonal disconnected covariance of Eqn.~\ref{eqn:cov} with $V = (1050\,h^{-1}\mathrm{Mpc})^3$ so as to replicate an analysis on an individual \texttt{Aemulus} test box. 
\par

\begin{figure}
    \centering
    \includegraphics[width=\columnwidth]{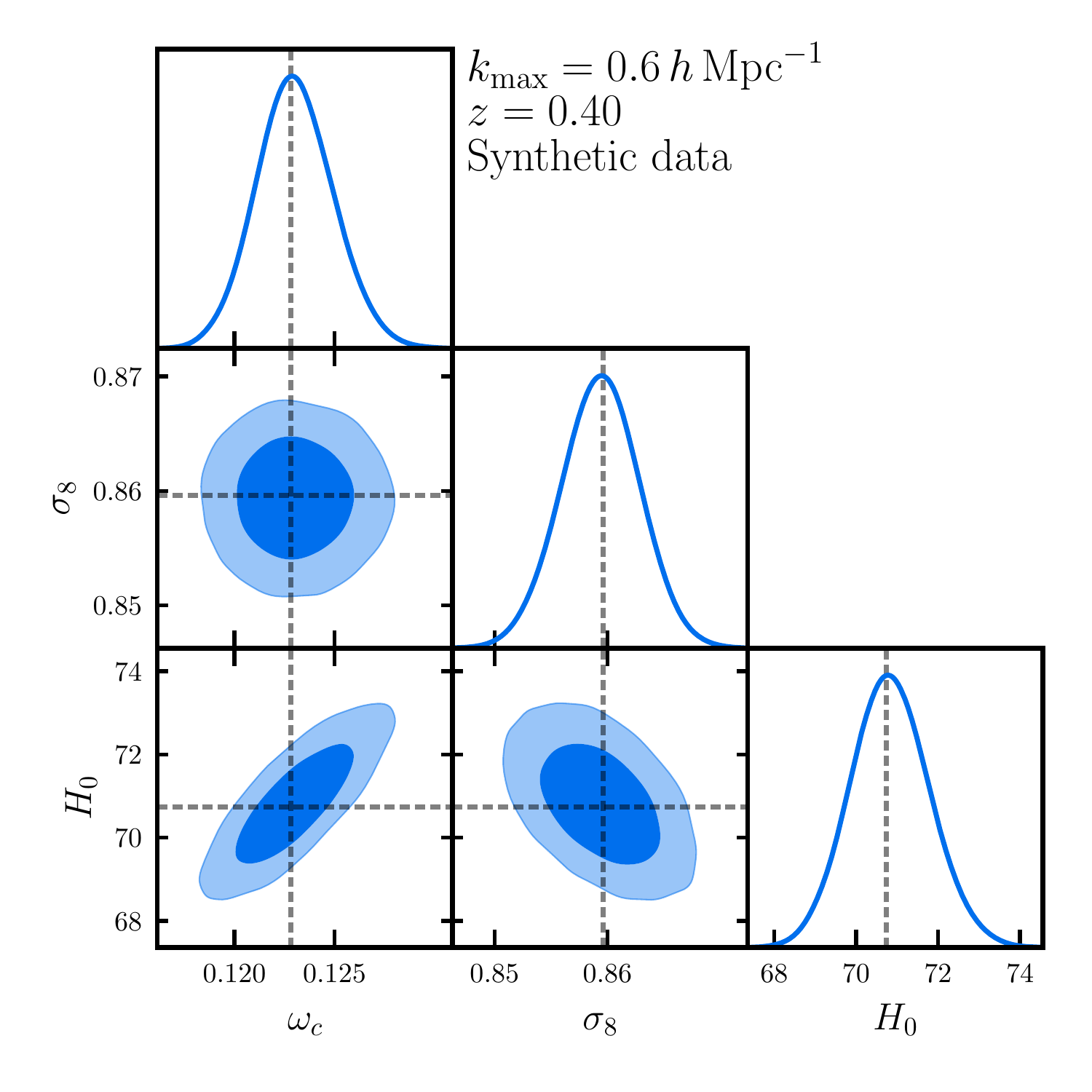}
    \caption{Cosmological parameter inference using the emulator where the data are a noiseless draw from itself. We vary the subset of parameters $\omega_c, \, \sigma_8,$ and $H_0$, using a Gaussian likelihood and purely disconnected covariance with volume $V = (1.05)^3 (h^{-1} \mathrm{Gpc})^3$. The fiducial values used to generate the data vector are shown in the dashed lines. The bias parameter posteriors are equally unbiased and Gaussian, but omitted from the figure for aesthetic purposes.}

    \label{fig:noiseless3param}
\end{figure}

The results of this first mock analysis are shown in in Fig.~\ref{fig:noiseless3param}. The three-parameter analysis constrains all cosmological and bias parameters in an unbiased fashion, indicating that there are no issues in fitting the emulator to itself at this volume. We also conduct seven-parameter analysis for $w$CDM parameters. The results returns unbiased posteriors relative to the true input values, however it is hard to constrain all $w$CDM parameters using a single halo sample at the volume of a single \texttt{Aemulus} box. For this reason, several of the cosmological parameters simply saturate the priors and remain unconstrained.

\subsubsection{Halo samples from the test suite}
\label{subsubsec:haloesample}

\begin{figure}
    \centering
    \includegraphics[width=\columnwidth]{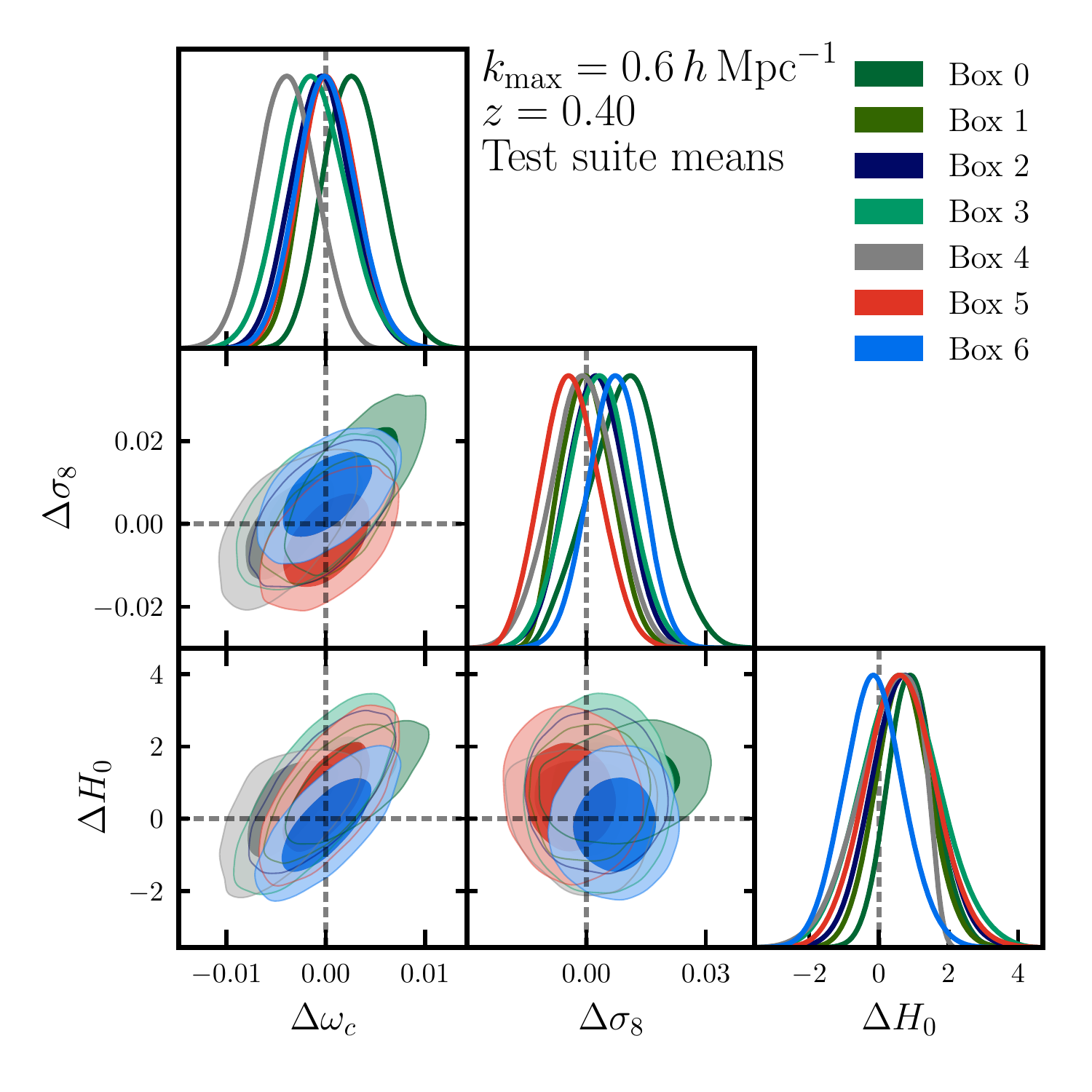}
    \caption{Cosmological parameter inference using the emulator fit to the mean of five realizations at the seven \texttt{Aemulus} test cosmologies. We vary the cosmological parameters $\omega_c, \, \sigma_8,$ and $H_0$, using a disconected covariance with volume $V = (1.05)^3 (h^{-1} \mathrm{Gpc})^3$, including a contribution arising from correlated emulator residuals. The contours are shown in the space of differences relative to the true cosmology of each box.}
    \label{fig:testsuite_multicosmo}
\end{figure}
A subsequent test we perform is inference on halo catalogs drawn from the \texttt{Aemulus} test suite. We refer to \texttt{Aemulus I} \citep{DeRose:2018xdj} for details on how the halo finding procedure was done. This fiducial halo sample contains the mass bin $13 \leq \log_{10} \left ( \frac{M_{h}}{h^{-1} M_\odot } \right) \leq 13.5$ at $z=0.4$. \par
We run a suite of chains for this halo sample, to assess emulator performance in terms of inferring cosmological parameters. We measure the halo--halo and halo--matter power spectra for each independent test box across the seven different test cosmologies. The data vector is averaged over the five independent realizations from the test suite. This set of chains allows us to assess the emulator biases such that they are less susceptible to projection effects. We can also study the cosmology dependence of the emulator in this way and the interplay between the bias coefficients of our model and cosmological parameters. \par

Using solely the purely disconnected covariance matrix in Eqn.~\ref{eqn:cov} leads to strong biases in inferred cosmological parameters, despite all residuals being smaller than 1 per cent as a function of scale. This can be understood by the fact that sample variance at small scales will eventually become smaller than the $1-2$ per cent emulator error observed in Fig.~\ref{fig:residuals} (see also Fig.~\ref{fig:emuerror_diag}).
However, the aforementioned figure allows us to estimate the emulator uncertainty as a function of scale. This can then be included as a separate contribution to the covariance matrix. We detail how this is done in Appendix \ref{appendix:cov} \footnote{All contour plots shown from this point forward will include the effects of emulator error unless stated otherwise.}.\par 
The result of the test is shown in Fig.~\ref{fig:testsuite_multicosmo}, with the full set of contours shown in Fig.~\ref{fig:testsuite_multicosmo_bigplot}. The cosmological parameters inferred scatter around the best fits for $\omega_c$ and $\sigma_8$, whereas they recover $H_0$ to within one standard deviation for most cosmologies but biased slightly high. However, we note these tests are conservative, as they neglect the contribution to the covariance matrix arising from shape noise, the lensing equivalent of shot-noise that would contribute to the $hm \, \times \, hm$ term of the covariance matrix. Given the conservative nature of this test we deem the emulator performance to be sufficient and continue with the final and most stringent test we consider in this work.
\subsubsection{A \redmagic sample from an independent simulation}
\label{subsubsec:redmagic}
\begin{table}
\centering
 \begin{tabular}{c c c c c c} 
 \hline\hline
 $\log M_{\mathrm{min}}$ & $\sigma_{\log M}$ & $f_c$ & $\log M_0$ & $\log M_1^{'}$ & $\alpha$ \\ [0.5ex] 
 12.1 & 0.4 & 0.13 & 11.45 & 13.73 & 1.48 \\ 
 \hline\hline
\end{tabular}
\caption{HOD parameters used to populate the \redmagic sample described in section \ref{subsubsec:redmagic}.}
\label{table:hod}
\end{table}
So far, we have reported tests performed on samples that originate either from the emulator itself or from the same suite of simulations used to construct it. It is also important that the model is useful for inference on spectra measured from tracer samples generated by independent methods, both in how halo samples are defined and the underlying $N$-body simulation used. For example, in \citet{modichenwhite19} it was shown that this hybrid Lagrangian bias model can successfully fit galaxy power spectra produced from a halo occupation distribution (HOD; see e.g.\ \citealt{Zheng:2004id}). \par 
We perform a final test: a simulated likelihood analysis with spectra produced from populating an independent $N$-body simulation with an HOD that matches the density and clustering properties of \redmagic galaxies \citep{Rozo:2015mmv}. \redmagic galaxies are the primary photometric Luminous Red Galaxy sample used in current and future weak lensing surveys \citep{Elvin-Poole:2017xsf}. \par 
The HOD parametrization we adopt is an extension of the model presented in \citet{Zheng:2007zg}, allowing for the central occupation at high mass to be less than unity
\begin{align}
    &\langle N_{\mathrm{cen}} (M) \rangle = \frac{f_c}{2} \left [ 1 + \mathrm{erf} \left ( \frac{\log M - \log M_{\mathrm{min}}}{\sigma_{\log M}} \right ) \right ], \\
    &\langle N_{\mathrm{sat}} (M) \rangle = \frac{1}{2} \left [1 +  \left (\frac{\log M - \log M_{\mathrm{min}}}{\sigma_{\log M}}  \right )\right]\left (  \frac{M - M_0 }{ M_1^{'} } \right )^\alpha .
\end{align}
The HOD parameters corresponding to the \redmagic samples used can be found in Table \ref{table:hod}, and are derived from a \redmagic sample selected from simulations similar to those presented in \citet{DeRose2019}. \par

We paint \redmagic galaxies onto halo catalogs measured from the \texttt{UNIT} simulations \citep{Chuang:2018ega} at $z\approx 0.59$, a redshift different from the \texttt{Aemulus} snapshots. A \texttt{UNIT} realization boasts a comparable volume to \texttt{Aemulus} of $V = 1\, (h^{-1}\mathrm{Gpc})^3$ at a significantly higher number of particles, $N = (4096)^3$. Every \texttt{UNIT} simulation has two realizations with opposite phases and fixed amplitudes. Averaging two-point statistics measured from these paired--fixed realizations leads to very high sample variance suppression at large scales, comparable to averaging $\sim 150$ simulations of the same volume. \par
\begin{figure*}
    \centering
    \includegraphics[width=\textwidth]{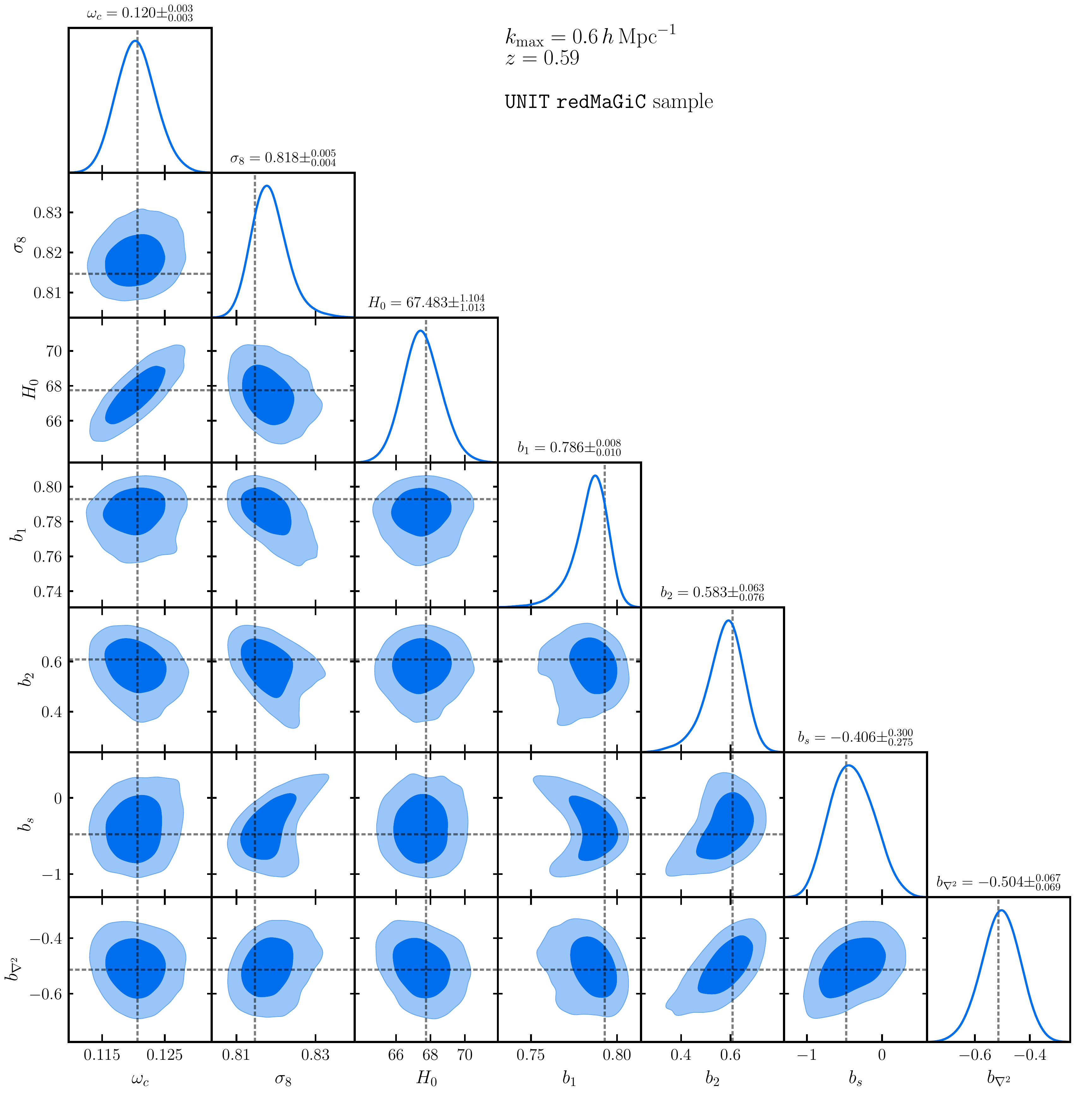}
    \caption{Cosmological parameter constraints from the \redmagic sample constructed from the \texttt{UNIT} simulations. The true cosmological parameters and the best-fit bias parameters assuming the true cosmology are shown in the dashed lines. All parameters are recovered to well within the one-sigma errors.}
    \label{fig:unitchain}
\end{figure*}

 
The cosmological parameter constraints corresponding to this test are shown in Fig.~\ref{fig:unitchain}. The emulator recovers the input cosmology of \texttt{UNIT} within its $68$ per cent contours. Although this test is idealized, the constraints inferred are promising if they translate even moderately well to a realistic analysis: a 2.5 per cent constraint on $\omega_c$, a 0.5 per cent constraint on $\sigma_8$ and a 1.6 per cent constraint of $H_0$. In a realistic lensing analysis one would expect these quantities to be degraded due to the inclusion of shape noise and only having access to two-dimensional lensing maps instead of the 3D matter field. Nevertheless, even a 100\% degradation of these constraints due to the aforementioned complications would still result in highly competitive measurements of these parameters. Note we adopt no priors beyond the (moderately informative) priors set by the boundaries of the \texttt{Aemulus} suite. \par 

\begin{figure}
    \centering
    \includegraphics[width=\columnwidth]{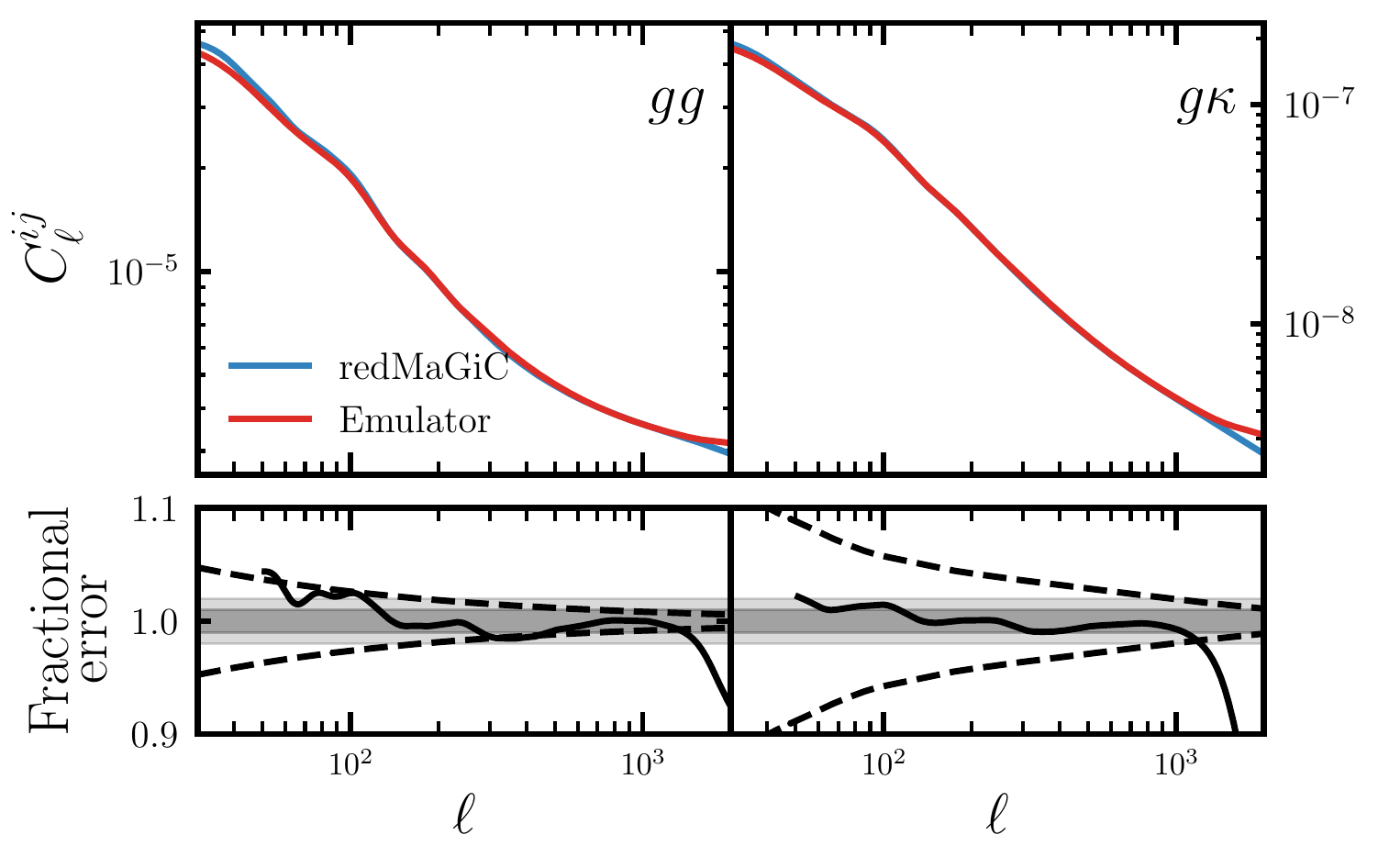}
    \caption{Residuals of the emulator fit to the \redmagic sample in the space of a projected analysis. Residuals are shown for the range $\ell \in [50, 2000]$. The redshift distributions of this analysis are consistent with those of current and upcoming surveys. The dashed envelope corresponds to the sample variance contribution in the absence of noise with sky coverage consistent with upcoming surveys and angular binning of $\Delta \ell = 50$. That is, shot/shape noise will only increase the size of this envelope. The light gray and dark gray bands correspond to 2 and 1 per cent error bands, respectively. }
    \label{fig:cell_lsst}
\end{figure}

The simulated likelihood analysis performed on this sample additionally allow us to quantify both model and emulator errors in a space that is closer to observations that will be carried out in the near future. As \redmagic galaxies are commonly used as lens samples in galaxy--galaxy lensing analyses, we can translate the $P^{hh}, P^{hm}$ residuals to those in the observables $C_\ell^{gg}, C_\ell^{g\kappa}$. We assume a redshift distribution $n(z)$ for \redmagic galaxies consistent with data \citep{Elvin-Poole:2017xsf} and fiducial parametrizations for the source sample that are consistent with those that will be achieved in future imaging surveys \citep{Mandelbaum:2018ouv}. For a \redmagic sample spanning $z=[0.45, 0.6]$ we present the results in Fig.~\ref{fig:cell_lsst}. The harmonic space observables are calculated assuming the Limber approximation, with the additional approximation that the residuals between 3D power spectra do not evolve as a function of redshift. The residuals stay within one per cent out to $\ell \approx 1000$.  If we instead use residuals from fitting the emulator at fixed cosmology to the same sample out to $k_{\rm max} = 1.0\, h {\rm Mpc}^{-1}$ the residuals remain within ten per cent out to $\ell_{\rm max} = 2000$, at the cost of worse performance at large scales. This indicates that the combined emulator and model error remain well under control for the analysis of current galaxy--galaxy lensing datasets.

\section{Conclusions}
In this work we have built an emulator to study the cosmology dependence of the model of \citet{modichenwhite19} for the two-point statistics of biased tracers. The model combines $N$-body simulations with a symmetries-based bias expansion to provide accurate predictions beyond the regime of validity of standard perturbative approaches. \par 
Specifically, we built an emulator for the cosmology and redshift dependence of the ten non-linear basis functions that span this model. We use measurements from the \texttt{Aemulus} suite of simulations, which has been designed to enable the construction of emulators that satisfy the modelling requirements of upcoming cosmic surveys. The model and emulation techniques used are general; there are no limitations to extending the range of validity given the availability of an improved suite of simulations. \par

We find that:
\begin{enumerate}
    \item The emulator recovers each basis spectrum to $\lesssim$ 1 per cent accuracy across a wide range of scales, $0.1 < k / \left(h^{-1}{\rm Mpc}\right) \leq 1.0$, and redshifts, $0 \leq z \leq 2$.
    \item The Lagrangian bias model is capable of capturing the clustering and lensing statistics of samples imbued with non-trivial amounts of secondary bias and contamination from baryonic physics.
    \item The test set used to validate the emulator can also be used to calibrate its `theoretical uncertainty'. This allows us to include contributions to the covariance matrix of an analysis related to model error, which cannot be neglected when pushing to small scales.
    \item The emulator, as constructed, can recover unbiased cosmological parameters from realistic simulated likelihood analyses. 
\end{enumerate}
These findings indicate that our emulator is a robust tool that can be readily applied to analyses of current and even upcoming datasets. The code will be made publicly available \href{https://github.com/kokron/anzu}{\faicon{github}} and can be integrated with modern sampling packages such as \href{https://github.com/CobayaSampler/cobaya}{\texttt{Cobaya}} \citep{torrado2020cobaya}. We also point out a few further directions to be investigated as a result of this work. \par

First, while the simulations used here are sufficient to obtain per cent level emulator accuracy, improved simulations will be important for maximizing the applicability of this model. The biggest immediate limitation of this emulator is the extent of the cosmological parameter space that it is trained on. We plan on running simulations over a broader parameter space, including massive neutrinos, in the near future. Another limiting factor in the current emulator construction is our ability to match the basis spectra measured from our simulations to their perturbation theory analogs at low $k$. Running larger simulation volumes, or implementing a method for sample variance mitigation such as that presented in \citet{chartier2020}, would ameliorate this issue by reducing noise in the $N$-body measurements. This will allow them to be matched more easily to the perturbation theory predictions at scales that are still safely within perturbative reach. Mismatches in the linear growth predictions from $N$-body simulations also limit the accuracy of the large scale matching. Simulations with more stringent time-stepping criteria would reduce these inaccuracies, at the cost of increased run-time. For this reason, methods that explicitly enforce linear growth on large scales may be worth exploring in the future \citep{Feng2016, Howlett2015}. Finally, the accuracy of the model for redshift evolution of the basis spectra in the current emulator is limited by the number of snapshots saved in the \texttt{Aemulus} suite. For this reason, saving snapshots with finer redshift resolution out to higher redshifts will be a priority when running future simulations to upgrade the current emulator. \par
While in this paper we have restricted ourselves to predictions of survey observables in Fourier space, one could use this same field-level approach to measure configuration-space correlation statistics instead. The model employed should also be able to describe the statistics of biased tracers at the field level, beyond two-point statistics. This includes both field-level characterizations of the Lagrangian bias model similarly to what was investigated in \citet{Schmittfull_2019} and higher order functions such as the bispectrum or the collapsed tri-spectra that form the connected component of covariance matrices. \par 
The field-level approach to bias modelling described in \citet{Schmittfull_2019} was recently extended to redshift space \citep{schmittfull2020modeling}.  For our emulator to be used to describe the statistics of 3D galaxy clustering in spectroscopic galaxy surveys, it would need to be extended to redshift space in a similar manner.  Alternatively, we note that the bias parameters in this model are equivalent to those of the Lagrangian perturbation theory of \citet{chen2020redshiftspace,Chen:2020fxs}.  This suggests one could perform a joint analysis that combines perturbation theory for describing the redshift-space clustering, where the 3D nature of the measurements allow tight constraints even on quasi-linear scales, and an emulator for describing projected statistics, which need to extend to smaller scales in order to beat down sample variance. In addition to providing a large dynamic range and sensitivity to both metric potentials, the combination of measurements would help to break bias parameter degeneracies and thus improve cosmological constraints.\par 
The second release of the \texttt{Aemulus} suite, \texttt{Aemulus-$\nu$}, will include two-fluid simulations that capture the effects of massive neutrinos on the matter density field. The techniques described in this paper can be translated to this new set of simulations to construct an emulator that can be used to constrain the sum of neutrino masses, one of the key science drivers of ongoing and future cosmological surveys. \par 
We leave these extensions to future work.

\section*{Acknowledgements}
We thank Simone Ferraro and An\v{z}e Slosar for helpful comments on a draft of the paper and Sean McLaughlin for many helpful discussions. We are grateful to the \texttt{Aemulus} collaboration for making the simulation suite used here publicly available. 
This work was supported in part by U.S. Department of Energy contracts to SLAC (DE-AC02-76SF00515) and by Stanford University. N.K.\ thanks the LSSTC Data Science Fellowship Program, which is funded by LSSTC, NSF Cybertraining Grant \#1829740, the Brinson Foundation, and the Moore Foundation.
S.C.\ is supported by the National Science Foundation Graduate Research Fellowship (Grant No.~DGE 1106400) and by the UC Berkeley Theoretical Astrophysics Center Astronomy and Astrophysics Graduate Fellowship.
M.W.\ is supported by the U.S. Department of Energy and the NSF. This research has made use of NASA's Astrophysics Data System and the arXiv preprint server. \par
Some of the computing for this project was performed on the Sherlock cluster. We would like to thank Stanford University and the Stanford Research Computing Center for providing computational resources and support that contributed to these research results.\par
Calculations and figures in this work have been made using \texttt{nbodykit} \citep{Hand_2018}, \texttt{GetDist} \citep{lewis2019getdist}, and the SciPy Stack \citep{2020NumPy-Array,2020SciPy-NMeth,4160265}.  
\section*{Data Availability}
The data underlying this article are available in the \href{https://aemulusproject.github.io/}{ \texttt{Aemulus} Project's} website.
\bibliography{main}
\bibliographystyle{mnras}

\appendix

\section{Including emulator error in the covariance matrix}
\label{appendix:cov}

As seen in Fig.~\ref{fig:residuals}, there is a scale-dependent error associated with our emulation scheme. This error is small, on the order of $\sim$ 1 per cent, and within the accuracy requirements for the next generation of surveys. However, at the smallest scales we would like to test this model, $k\simeq 0.6\,h\,\mathrm{Mpc}^{-1}$, it will often be larger than the combined cosmic variance and shot noise (and absence of shape noise) of our tests. In this regime, the combination of using the average of only five boxes as our data, the approximate disconnected form of the covariance in Eqn.~\ref{eqn:cov} and failing to include model uncertainty in an analysis could then lead to biased inference on cosmological parameters \citep{Baldauf:2016sjb,Chudaykin:2020hbf}. \par 
Since the \texttt{Aemulus} test suite is composed of 35 simulations, at seven distinct points of cosmological parameter space, we can use the emulator residuals at these points to construct a model for the theoretical uncertainty. In this appendix we discuss our procedure to construct this model and study its impact when employed in inference. \par 
Let $\bar{P}_{XY} (k, \Omega_i)$ be the mean basis spectrum measured from five \texttt{Aemulus} boxes at the cosmology $\Omega_i$. For a given box, we define normalized emulator residuals as 
\begin{equation}
    \hat{r}^{XY} (k) = \frac{\hat{P}_{XY}(k, \Omega_i) - P^{\mathrm{Emu}}_{XY} (k, \Omega_i))}{\bar{P}_{XY} (k, \Omega_i)},
\end{equation}
where $P^{\mathrm{Emu}}_{XY}$ is the emulator prediction at the same cosmology. Normalized this way, we assume the residuals are cosmology independent. At each redshift we have 35 sets of residuals. With these measurements we can build an estimate of the residual correlation matrix
\begin{equation}
    \label{eqn:emucorr}
    \mathrm{Corr}^{\mathrm{Emu}}(k, k') = \frac{\mathrm{Cov}[\hat{r}^{XY}(k), \hat{r}^{XY}(k')]}{\sqrt{\mathrm{Cov}(k, k)\mathrm{Cov}(k', k')}}
\end{equation}
which captures how correlated the emulator residuals are across the test set as a function of scale. The quantities in the numerator and denominator of Eqn.~\ref{eqn:emucorr} are the same, but we apply the shorthand $\mathrm{Cov}(k, k) \equiv \mathrm{Cov}[\hat{r}^{XY}(k), \hat{r}^{XY}(k)] $ to not overload the expression.
We proceed to define an emulator floor, $f_{\mathrm{Emu}}$, specifying what fraction of the signal is of the order emulator error. From Fig.~\ref{fig:residuals}, the dominant source of uncertainty will come from the error in the $P_{11}$ spectrum. This implies $f_\mathrm{Emu} \simeq 0.01$ at small scales for redshifts $z>0$. We then estimate that the emulator error will scale as
\begin{equation}
    \mathrm{Cov}^{\mathrm{Err}}(k, k') = (f_{\mathrm{Emu}}P_{hh,hm}(k))^2 \times \mathrm{Corr}^{\mathrm{Emu}}(k, k'),
\end{equation}
where $P_{hh, hm}$ is used depending on whether we are including this contribution to the block corresponding to the halo--halo correlation or the halo-matter correlation. We then add this contribution in quadrature to Eq.~\ref{eqn:cov}
\begin{equation}
    \label{eqn:fullcov}
    \mathrm{Cov}(k, k') = \mathrm{Cov}^G(k, k') + \mathrm{Cov}^{\mathrm{Err}}(k, k') .
\end{equation}
We run chains with the covariance in Eq.~\ref{eqn:fullcov}, as well as chains including only the diagonal contribution due to uncertainty, which we will call the `floor' covariance. The contours for cosmological parameters are shown in Fig.~\ref{fig:chain_covtest}. While this is clearly an approximate treatment, we observe that including this contribution helps prevent significant biases in cosmological parameter inference due to the noisy input data and very low noise assumed in the fit.

\begin{figure}
    \centering
    \includegraphics[width=1\columnwidth]{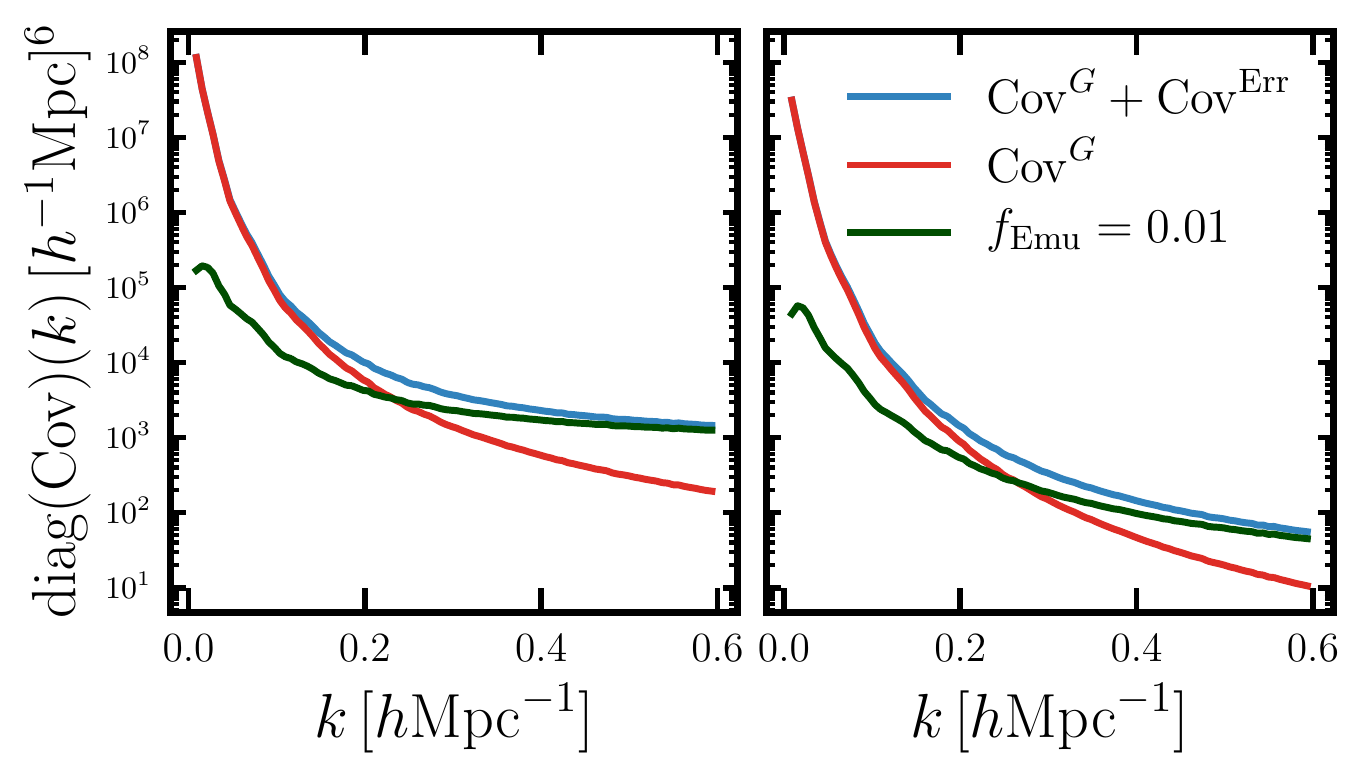}
    \caption{Comparison of our model for emulator uncertainty compared to the disconnected component of the covariance matrix. The left panel corresponds to $P_{hh}P_{hh}$ contribution and the right panel to $P_{hm}P_{hm}$. 
    }
    \label{fig:emuerror_diag}
\end{figure}

\begin{figure}
    \centering
    \includegraphics[width=\columnwidth]{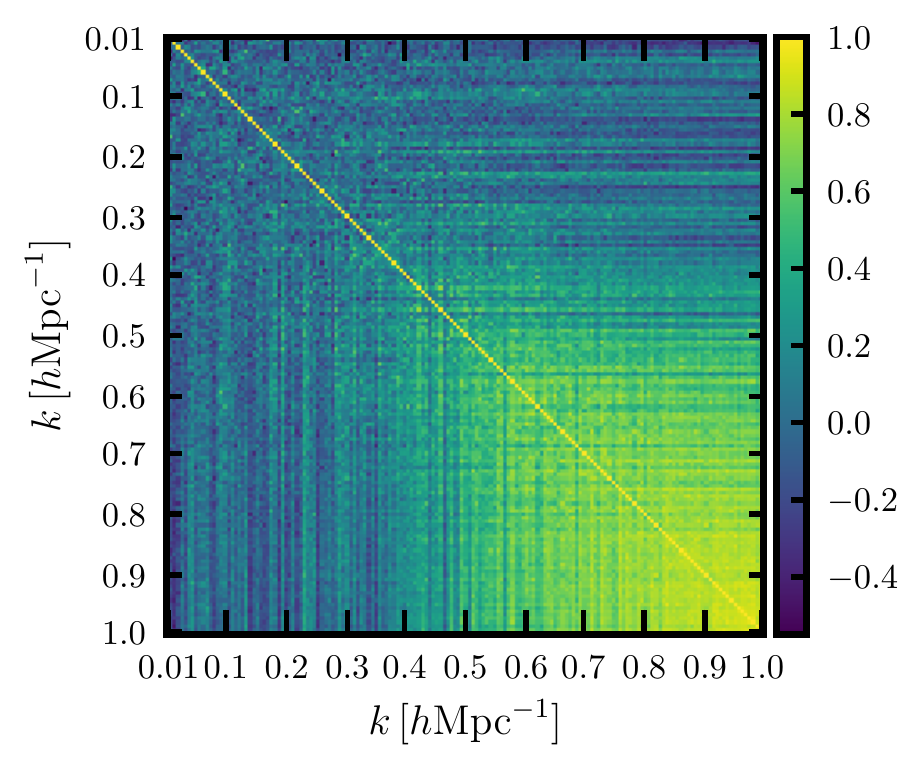}
    \caption{Correlation matrix of emulator residuals described in Appendix \ref{appendix:cov}. We see at small scales, past $k \simeq 0.4 h {\rm Mpc}^{-1}$, the emulator residuals are significantly correlated.}
    \label{fig:corrmat}
\end{figure}
\begin{figure}
    \centering
    \includegraphics[width=\columnwidth]{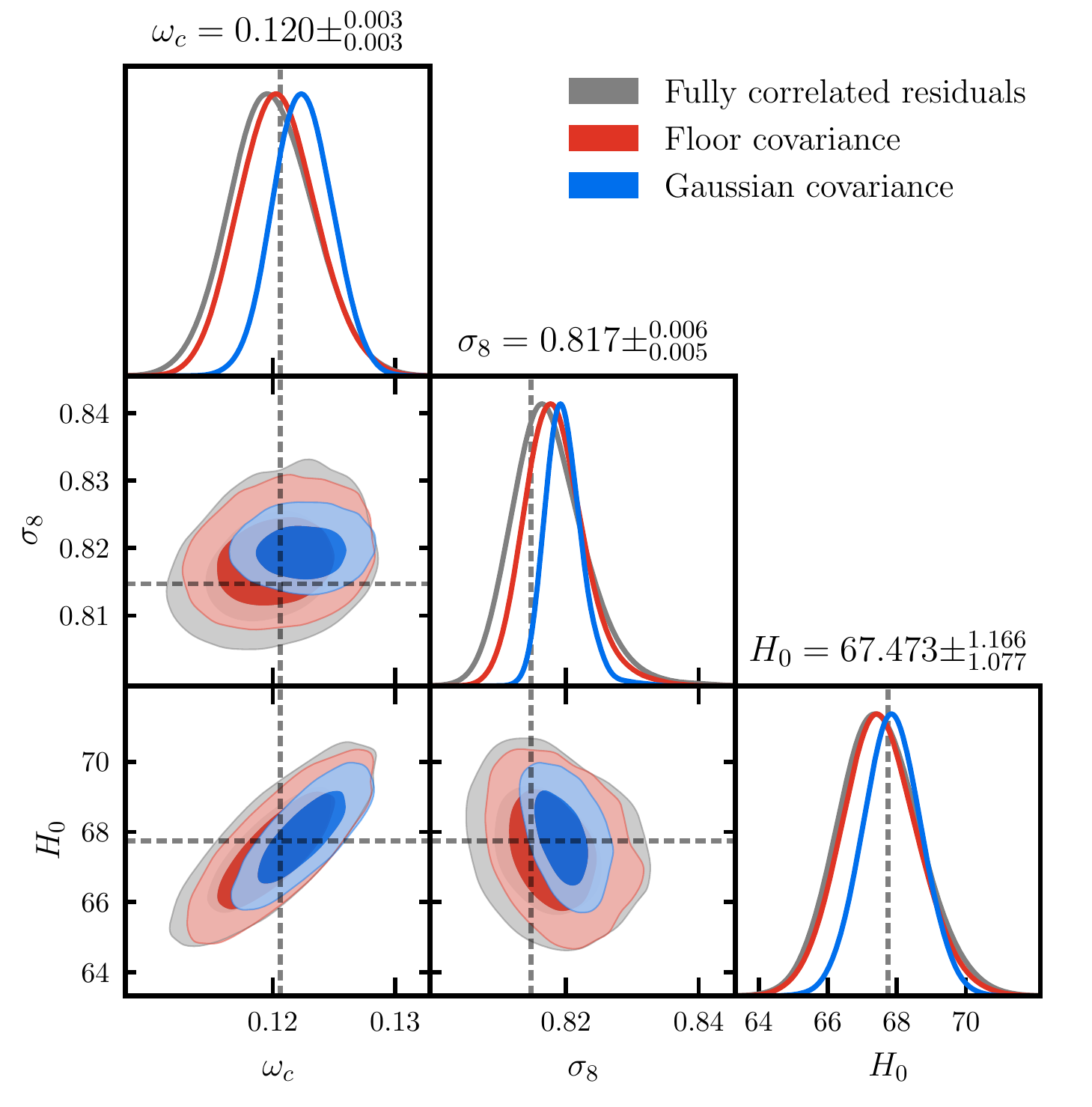}
    \caption{\texttt{UNIT} contours with the different covariance forms discussed in Appendix \ref{appendix:cov}. Chains are run with the standard scale cuts of $k_{\rm max} = 0.6\,  h^{-1} {\rm Mpc}$.}
    \label{fig:chain_covtest}
\end{figure}

\section{Subsets of the bias model}

\label{appendix:biassubset}
A common critique of EFT-based models is that they are over-parametrized, and can fit to any signal due to the large number of free parameters. For perturbative Lagrangian bias models, this question has been previously explored in the context of CMB lensing cross-correlations. In \cite{2017JCAP...08..009M}, it was shown that significant biases are obtained in $\sigma_8$ in these analyses if one uses a simplified model with linear galaxy bias and non-linear matter power spectra. To address whether this holds for our model, we run a series of tests of the emulator, with differing subsets of the bias parameters set to zero. The full set we adopt is
\begin{enumerate}
    \item `All $b_i$'s ', the full bias parametrization.
    \item `$b_1$ only', where $b_2 = b_{s^2} = b_{\nabla^2} = 0$.
    \item `$b_1, \, b_{\nabla^2}$', where $b_2 = b_{s^2} = 0$.
    \item `No $b_{s^2}$', where $b_{s^2} = 0$.
    \item `No $b_{2}$', where $b_2 = 0$.
\end{enumerate}

\begin{figure*}
    \centering
    \includegraphics[width=1.5\columnwidth]{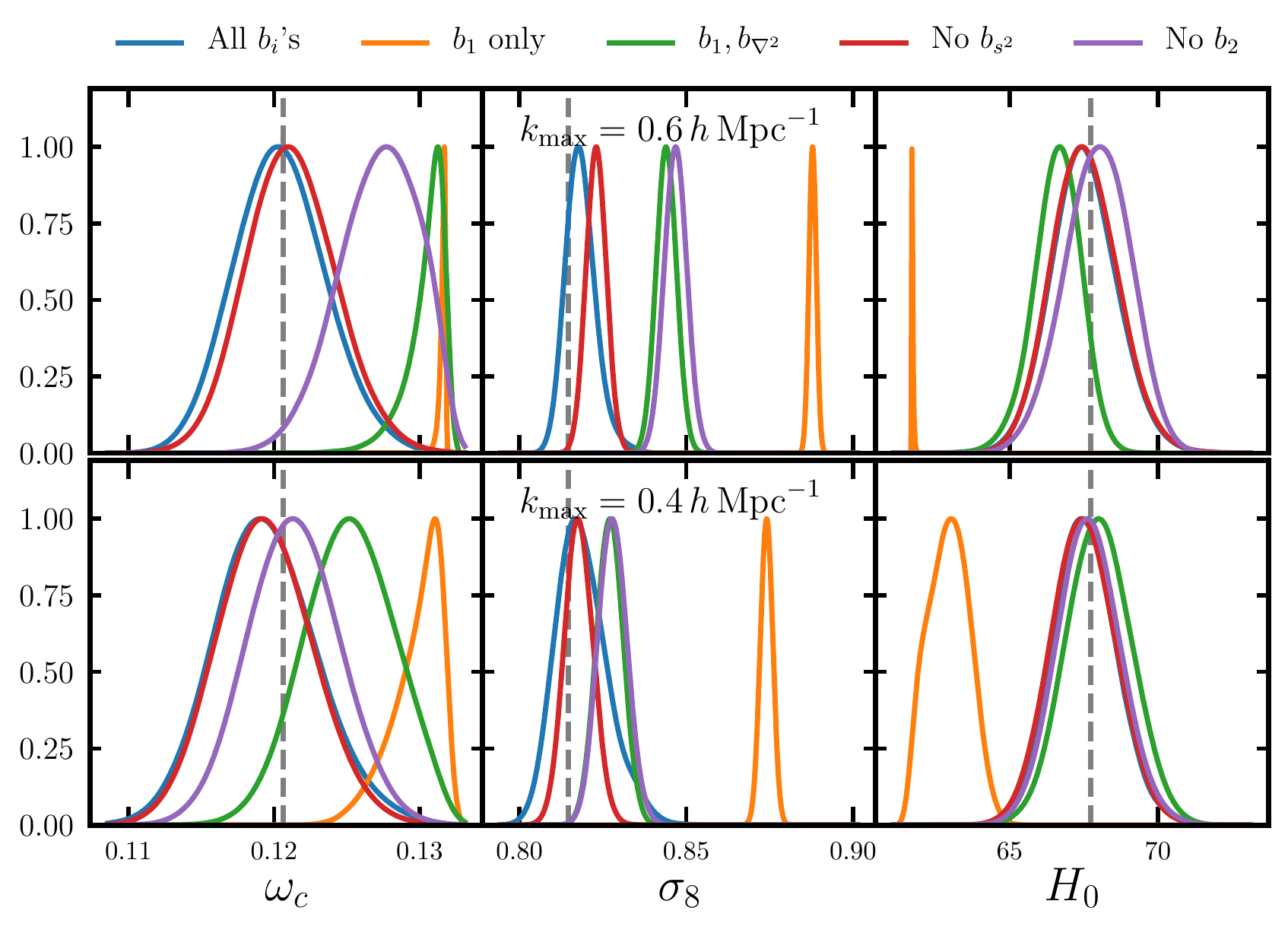}
    \caption{Posteriors for varying subsets of the bias model in Eqn.~\ref{eqn:lagbias}, for two different scale cut configurations.}
    \label{fig:biassubset}
\end{figure*}
A contribution due to shot-noise is included in all of the chains. All chains in Fig.\  \ref{fig:biassubset} are run with the same data vector and covariance matrices, and the $k_{\rm max}$ cuts highlighted in each row. We observe significant biases for every subset of bias parameters, except for the complete parameterization which recovers the input cosmological parameters as previously discussed in section \ref{subsubsec:redmagic}. This implies, at least in this simplified analysis, that the full set of bias parameters is required to achieve unbiased inference with this model. \par
To check the scale-dependence of the importance of the full parameterization, the second row of  Fig.~\ref{fig:biassubset} repeats this test limiting ourselves to $k_{\mathrm{max}} = 0.4 h {\rm Mpc}^{-1}$. The full bias model and the subset including only linear, quadratic and higher derivative biases perform comparatively well. \par 

\section{The $k \to 0$ limit of the emulator}
\label{appendix:lptrecover}
In this appendix we investigate the impact of not correctly recovering large-scale linear growth in $N$-body simulations on the emulator, as highlighted in \S\ref{fig:fieldlevel}. We implement two different forms of enforcing consistency with linear theory at large scales:
\begin{itemize}
    \item Strictly reverting to LPT at $k < k_{\rm min}$. This introduces a `kink' in the basis spectra predicted by the emulator. 
    \item Extrapolating the principal component predictions out to $k < k_{\rm min}$, but with a filter to enforce linear growth.
\end{itemize}
The filter is applied to the $\Gamma^{XY} (k)$ that we use to build the emulator, 
\begin{align}
\Gamma^{XY} (k, {\bf \Omega)} \to F(k) \Gamma^{XY} (k, {\bf \Omega}).
\end{align}
With this filtering approach, we recover LPT at large scales by construction without the discontinuity introduced by simply forcing LPT after some transition. The functional form we adopted for $F(k)$ is
\begin{align}
    F(k) = \frac{1}{2} \left [ 1 + \tanh \left ( \alpha \frac{k - k_*}{k_*} \right )\right ].
\end{align}
This quantity asymptotes to 0 at large scales, ensuring the $\Gamma^{XY}$ are 0, and thus the ratios are consistent with unity. Fiducial values adopted are $k_* = 0.125$ and $\alpha =2.5$ but the impact is similar for other values. \par
Since the samples we use to test the emulator are also derived from boxes with incorrect growth, for all figures in this paper we adopt a `fiducial model' where we use the $\Gamma^{XY}$ with no corrections at large-scales. The emulator then has large-scale growth compatible with the boxes. \par 
If we perform a simulated likelihood analysis with the other variants that enforce LPT at large scales we see small shifts in some cosmological parameters away from their true values. The shifts in parameters are all less than one $\sigma$, and one must keep in mind that the noise levels in our analysis are quite stringent (for example, we have no shape noise in the simulated lensing constraint). When phrased in terms of the uncertainties in parameters obtained by recent analyses \citep{2020arXiv200715632H}, these shifts are less than $(1/4)\,\sigma$. 
\begin{figure*}
    \centering
    \includegraphics[width=\textwidth]{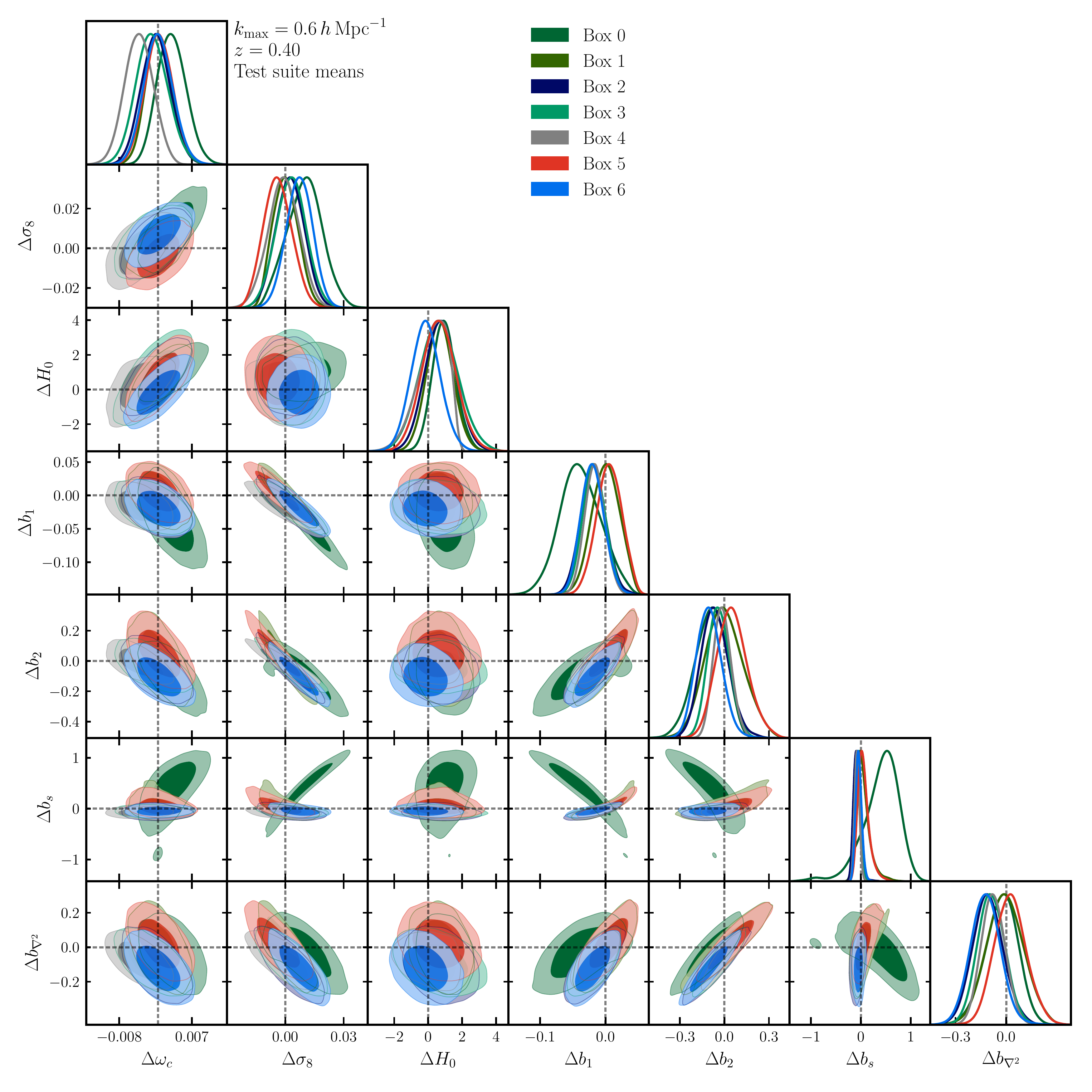}
    \caption{The same chains as Fig.~\ref{fig:testsuite_multicosmo} but showing all parameters varied.}
    \label{fig:testsuite_multicosmo_bigplot}
\end{figure*}
\end{document}